\documentclass[12pt]{article}

\usepackage{graphicx}
\usepackage{amsfonts}
\usepackage{amssymb}
\usepackage[mathscr]{eucal}
\usepackage{amsthm,amsmath,bm,color,ctable}
\usepackage{enumerate,natbib}
\usepackage{dsfont, multirow}
\usepackage{epsfig}
\usepackage{verbatim}
\usepackage{epstopdf}
\usepackage{natbib}
\graphicspath{{figures/}}

\makeatletter

\newcommand{\Rmnum}[1]{\expandafter\@slowromancap\romannumeral #1@}
\makeatother
\newcommand{\mc}[2]{\multicolumn{#1}{c}{#2}}

\newcommand{\Var}{\mathop{\mbox{\sf Var}}}

\newcommand{\E}{\mathop{\mbox{\sf E}}}     

\newcommand{\bbeta}{\boldsymbol{\beta}}
\newcommand{\balpha}{\boldsymbol{\alpha}}
\newcommand{\bvarepsilon}{\boldsymbol{\varepsilon}}

\renewcommand{\theequation}{\thesection\hspace{+0.25cm}\arabic{equation}}

\newtheorem{thm}{\hspace{1mm}Theorem}
\newtheorem{exm}{\hspace{1mm}Example}
\newtheorem{cor}{\hspace{1mm}Corollary}

\newcommand{\etal}{{\sl et al. }}
\newcommand{\bepsilon}{\mbox{\boldmath$\epsilon$}}
\newcommand{\bY}{\mbox{\bf Y}}

\newcommand{\bZ}{\mbox{\bf Z}}
\newcommand{\bM}{\mbox{\bf M}}
\newcommand{\bB}{\mbox{\bf B}}

\begin{document}

\title{\Large\bf Estimation of Partially
Linear Regression  Model under Partial Consistency Property}

\author{\large Xia Cui,\thanks{School of Mathematics and Information Science, Guangzhou University, Guangzhou, China } \quad \large Ying Lu\thanks{Center for the Promotion of Research Involving Innovative Statistical Methodology, Steinhardt School of Culture,
Education and Human Development,New York
University, New York, USA}  \quad and \quad
{\large Heng Peng }\thanks{Department of Mathematics, Hong Kong
Baptist University, Hong Kong} }

\date{}
\maketitle

\begin{abstract}\small In this paper, utilizing recent theoretical results
in high dimensional statistical modeling, we propose a model-free yet computationally simple approach to estimate
the partially linear model $Y=X\beta+g(Z)+\varepsilon$. Motivated by the partial consistency phenomena, we propose to model $g(Z)$ via
incidental parameters. Based on partitioning the support of $Z$, a simple local average is used to
estimate the response surface. The proposed method seeks to strike a balance between computation
burden and efficiency of the estimators while minimizing model bias. Computationally this approach only
involves least squares.  We show that given the inconsistent estimator of $g(Z)$, a root $n$
consistent estimator of parametric component $\beta$ of the partially linear model can be obtained with little
cost in efficiency.  Moreover, conditional on the $\beta$ estimates, an optimal estimator of $g(Z)$ can then
be obtained using classic nonparametric methods.  The statistical inference problem regarding
$\beta$ and a two-population nonparametric testing problem regarding $g(Z)$ are considered.
Our results show that the behavior of test statistics are satisfactory. To assess the performance of our method
in comparison with other methods, three simulation studies are conducted and a real dataset about risk factors of birth
weights is analyzed.

\end{abstract}

\


\newpage

\section{Introduction}
\setcounter{equation}{0}

In statistics, regression analysis is a family of important techniques that estimate the relationship
between a continuous response variable $Y$ and covariates $X$ with dimension $p$, $Y=f(X)+\epsilon$.
Parametric regression models specify the regression function in terms of a small number of parameters.
For example, in linear regression, a linear response surface $E(Y)=X\beta$ is assumed and determined
by $p \times 1$ vector of $\beta$. The parametric methods are easy to estimate and are widely used in
statistical practice as parameters $\beta$ can be naturally interpreted as the ``effects of X on Y''.
However the requirement of a pre-determined functional form can increase the risk of model misspecification,
which leads to invalid estimates. In contrast, nonparametric methods assume no predetermined
functional form and $f(X)$ is estimated entirely using the information from the data. Various kernel
methods or smoothing techniques have been developed to estimate $f(X)$. In general, these methods use local
information about $f(X)$ to blur the influence of noise at each data point. The bandwidth, $h$, determines
the width of the local neighborhood and the kernel function determines the contribution of data points in
the neighborhood. The bandwidth $h$ is essential to the nonparametric estimator $\hat{f}(X)$. Smoother
estimates of $f(X)$ are produced as $h$ increases and vice versa. As a special case, the local linear model
reduces to linear regression when $h$ spans the entire data set with a flat kernel. The choice of $h$ is data
driven and can be computationally demanding as the dimension of $X$ increases. Moreover, nonparametric
estimation suffers the curse of dimensionality which requires the sample size to increase exponentially with
the dimension of $X$. In addition, most kernel functions are designed for continuous variables and it is not
natural to incorporate categorical predictors. Hence a fully nonparametric approach is rarely useful to estimate
the regression function with multiple covariates.

The partially linear model, one of the most commonly used semi-parametric regression models,
\begin{equation}\label{PL_model}
Y_i=X_i^\top\bbeta+g(Z_i)+ \varepsilon_i,\quad i=1,\ldots,n
\end{equation}
offer an appealing alternative in that it allows both parametric and nonparametric specifications in the
regression function. In this model, the covariates are separated into parametric components $X_i=(X_{i1},
\ldots, X_{ip})^\top$ and nonparametric components $Z_i=(Z_{i1}, \ldots, Z_{iq})^\top$.  The parametric part of the model can be
interpreted as a linear model, while the nonparametric part frees the model from stringent structural
assumptions. As a result, the estimates of $\beta$ are also less affected by model bias. This model has gained
great popularity since it was first introduced by \citet*{Engle&Granger&Rice&Weiss:1986} and has been widely applied
in economics, social and biological sciences. A lot of work have also been devoted to the estimation of the partially
linear models. \citet*{Engle&Granger&Rice&Weiss:1986} and many others study the penalized least squares method
for partially linear regression models estimation. \citet*{Robinson:1988} introduces a profile least squares
estimator for $\bbeta$ based on the Nadaraya-Watson kernel estimate of the unknown function $g(\cdot)$. \citet*{Heckman:1986},
\citet*{Rice:1986}, \citet*{Chen:1988} and \citet*{Speckman:1988} study the consistency properties of the estimate
of $\bbeta$ under different assumptions. \citet*{Schick:1996} and \citet*{Liang&Hardle:1997} extend the root
$n$ consistency and asymptotic results for the case of heteroscedasticity. For models with only specification
of the first two moments, \citet*{Severini&Staniswalis:1994} propose a quasi-likelihood estimation method.
\citet*{Hardle&Mammen&Muller:1998} investigate nonparametric testing problem of the unknown function $g(\cdot)$.
Among others, \citet*{Hardle&Liang&Gao:2000} provide a good comprehensive reference of the partially linear model.

Most of the above methods are based on the idea of first taking the conditional expectation give $Z_i$
and then subtracting the conditional expectations in both sides
of (\ref{PL_model}). This way, the function $g(\cdot)$
 disappears,
\begin{equation}\label{tra-model}
Y_i-\E(Y_i|Z_i) = \{X_i-\E(X_i|Z_i)\}^{\top}\bbeta + \varepsilon_i,
\ i=1,\ldots,n.
\end{equation}
If the conditional expectations were known, $\bbeta$ could be readily estimated via regression techniques. In practice,
the quantities $\E(Y|Z)$ and $\E(X|Z)$ are estimated via nonparametric method. The estimation of these conditional
expectations is very difficult when the dimension of $Z_i$ is high. Without accurate and stable estimates of those
conditional expectations, the estimates of $\bbeta$ will also be negatively affected. In fact, \citet*{Robinson:1988},
\citet*{Andrews:1994} and \citet*{LiQi:1996} obtain the root-$n$ consistency of the estimator of $\bbeta$ under an important
bandwidth condition with respect to the nonparametric part: $\sqrt{n}\Big(h^4+\frac{1}{nh^q}\Big)\rightarrow 0$.
Clearly, this condition breaks down when $q>3$.

To circumvent the curse of dimensionality, $g(Z)$ is often specified in terms of additive structure of one-dimensional
nonparametric functions, $\sum_{j=1}^q g_j(Z_j)$. This is the so-called generalized additive model. In theory, if the specified
additive structure corresponds to the underlying true model, every $g_j(\cdot)$ can be estimated with desired one-dimensional
nonparametric precision, and $\bbeta$ can be estimated efficiently with optimal convergent rate. But in practice, estimating
multiple nonparametric functions is related to complicated bandwidth selection procedures, which increases computation complexity
and makes the results unstable. Moreover, when variables $\{Z_j\}$ are highly correlated, the stability and accuracy of such
additive structure in partially linear regression model is problematic \citep*[see][]{Jiang&Fan&Fan:2010}. Lastly, if the
additive structure is misspecified, for example, when there are interactions between the nonparametric predictors $Z$, the model
and the estimation of $\bbeta$ will be biased.

In this paper, we propose a simple least squares based method to estimate the parametric component of model
(\ref{PL_model}) without complicated nonparametric estimation. The basic idea is as follows. Since the value
of $g(Z)$ at each point is only related to the local properties of $g(\cdot)$, it can be represented by a set
of incidental parameters that are only related to finite local sample points. Inspired by the partial consistency
property (Neyman and Scott, 1942; \cite{Lancaster:2000}; \cite{Fan&Peng&Huang:2005}), we propose to approximate
$g(Z)$ using local averages over small partitions of the support of $g(\cdot)$. The parametric parameters $\bbeta$
can then be estimated using profile least squares. Following the classic results about the partial consistency property
(\cite{Fan&Peng&Huang:2005}), we show that, under moderate conditions this estimator of $\bbeta$ has optimal
root-$n$ consistency and is almost efficient. Moreover, given a good estimate of $\bbeta$, an improved estimate of the
nonparametric component $g(Z)$ can be obtained. Compared to the classic nonparametric approach, this method is not only
easy to compute, it also readily incorporates covariates $Z$ when they contain both continuous and categorical variables.
We also explore the statistical inference problems regarding the parametric and nonparametric components under the proposed
estimating method. Two test statistics are proposed and their limiting distributions are examined.

The rest of the paper is organized as follows. In section~2, followed by a brief technical review of the partial
consistency property, we propose a new estimation method of the parametric component for the partially linear
regression model. The consistency of the parameter estimates are shown when the nonparametric component consists of
univariate, one continuous and one categorical variable or two highly correlated continuous variables.  The inference
methods of the partially linear regression model are discussed in Section 3. Numerical studies assessing the performance
of the proposed method in comparison with existing alternative methods are presented in Section~4. A real data example
is analyzed in Section~5. In Section~6 we offer an in-depth discussion about the implications of the proposed method and
further directions. Technical proofs are relegated to the Appendix.


\section{Estimating partially linear regression model under partial consistency property}
\setcounter{equation}{0}

\subsection{Review of partial consistent phenomenon}
The partial consistency property refers to a phenomenon when a statistical model contains nuisance parameters
whose number grows with sample size; although the nuisance parameter themselves cannot be estimated consistently,
the rest of the parameters sometimes can be. Neyman and Scott (1942) first studied this phenomenon. Using their
terminology,  the nuisance parameters are ``incidental" since each of them is only related to finite sample points,
and the parameters that can be estimated consistently are ``structural" because every sample point contains
their information. The partial consistency phenomenon appears in mixed effect models, models for longitudinal data, and
panel data in econometrics, see \cite*{Lancaster:2000} etc. In one JASA discussion paper, \cite*{Fan&Peng&Huang:2005}
formally studied the theoretical properties of parameter estimators under partial consistency and their applications
to microarray normalization. They consider a very general form of regression model
\begin{equation} \label{Fan_Model}
\bY_n =\bB_n \balpha_n +\bZ_n \bbeta +\bM+\bepsilon_n, \quad n =J\times I,
\end{equation}
where $\bY_n=(Y_1,\ldots,Y_n)^T$, $\bB_n=\mathbf{I}_J \otimes \mathbf{1}_I$  is an $n \times J$ design matrix, $I$
is assumed to be a constant and $J$ grows with sample size $n$. $\bZ_n$ is an $n\times d$ random matrix with $d$ being
the dimension of $\bbeta$, $\bM=(m(X_1),\ldots,m(X_n))$ is an nonparametric function, and
$\bepsilon_n=(\varepsilon_1,\ldots,\varepsilon_n)$ is a vector of i.i.d. errors. In the above model, $\balpha_n$ is a
vector of incidental parameters as its dimension $J$ increases with sample size, $\bbeta$ and $\bM$ are the structural
parameters.  Fan \etal (2005)  show that $\bbeta$ and $M$ can be estimated consistently and even nearly efficient when
the value of $I$ is moderately large,
$$
\sqrt{n}(\hat{\bbeta}-\bbeta) \sim
\mathcal{N}(0,\frac{I}{I-1}\sigma^2\Sigma^{-1}),
$$
the factor $I/(I-1)$ is the price to pay for estimating the nuisance parameters $\balpha_n$.

\subsection{Estimating partially linear model under partial consistency}

First we apply our proposed strategy to estimate a partially linear regression model with one-dimensional
nonparametric component,
\begin{equation}\label{SPL}
Y_i=X_i\bbeta+g(Z_i)+\varepsilon_i, \ i=1,\ldots,n,
\end{equation}
where $g(\cdot)$ is an unknown function, $Z_i \in R^1$ is a continuous random variable, and other
assumptions for the model are similar as those imposed on the model (\ref{Fan_Model}).
Without loss of generality, we assume that $Z_i$ are i.i.d random variables and follow $[0,1]$ uniform
distribution, and is sorted as $0 \le Z_1 \le Z_2 \ldots \le Z_n \le 1$ based on their realized values.
Then we can partition the support of $Z_i$ into $J=n/I$ sub-intervals such that the $j$th interval
covers $I$ different random variables with closely realized values from $z_{(j-1)I+1}$ to $z_{jI}$.
If the density of $Z_i$ is smooth enough, these sub-intervals should be narrow and the values of
$g(\cdot)$ over the same sub-interval should be close and
$g(Z_{(j-1)I+1})\approx g(Z_{(j-1)I+2}) \cdots \approx g(Z_{jI})\approx \alpha_j$
where $\alpha_j=\frac{1}{I}\sum_{i=1}^I g(Z_{(j-1)I+i})$. Then the nonparametric part of model
(\ref{SPL}) can be reformulated in terms of partially consistent observations and rewritten in the
form of the model (\ref{Fan_Model})
\begin{equation} \label{PLinear_Model}
\mathbf{Y}_n=\mathbf{B}_n\balpha_n+\mathbf{X}_n\bbeta+\bvarepsilon^\ast_n,
\quad n=J\times I
\end{equation}
with $\varepsilon^\ast_{(j-1)I+i}=\varepsilon_{(j-1)I+i}+g(Z_{(j-1)I+i})-\frac{1}{I}\sum_{i=1}^I g(Z_{(j-1)I+i}).$
It is easy to see that the second term in $\varepsilon^\ast_{(j-1)I+i}$ is the approximation error.
Normally when $I$ is a small constant, it is of order $O(1/J)$ or $O(1/n)$ , and much smaller than
$\varepsilon$. Hence the approximation error can be ignored and it is expected that, similar to as in
(\ref{Fan_Model}),  $\bbeta$ in the model (\ref{SPL}) or (\ref{PLinear_Model}) can be estimated almost
efficiently even when $g(\cdot)$ in (\ref{SPL}) is not estimated consistently.

Model (\ref{PLinear_Model}) can be easily estimated by profile least squares,
\begin{equation}
\sum\limits_{j=1}^J\sum\limits_{i=1}^{I}(Y_{(j-1)I+i} -
X_{(j-1)I+i}\,\bbeta - \alpha_j)^2.
\end{equation}
the estimates of $\bbeta$ and $\alpha_j$ can be expressed as follows,
\begin{equation}\label{est}
\left\{
\begin{split}
& \hat{\bbeta} =
\Big\{\sum\limits_{j=1}^J\sum\limits_{i=1}^{I}\{X_{(j-1)I+i}-\frac{1}{I}\sum\limits_{i=1}^{I}X_{(j-1)I+i}\}^T
\{X_{(j-1)I+i}-\frac{1}{I}\sum\limits_{i=1}^{I}X_{(j-1)I+i}\}\Big\}^{-1}\\
&\hspace{1cm} \times
\Big\{\sum\limits_{j=1}^J\sum\limits_{j=1}^{J}\{X_{(j-1)I+i}-\frac{1}{I}\sum\limits_{i=1}^{I}X_{(j-1)I+i}\}^T
\{Y_{(j-1)I+i}-\frac{1}{I}\sum\limits_{i=1}^{I}Y_{(j-1)I+i}\}\Big\},\\
& \hat{\alpha}_j =
\frac{1}{I}\sum\limits_{i=1}^{I}\left\{Y_{(j-1)I+i}-X_{(j-1)I+i}\hat{\bbeta}\right\}.
\end{split}
\right.
\end{equation}

We have the following theorem for the above profile least squares estimate of $\bbeta$ under the model
(\ref{SPL}) or (\ref{PLinear_Model}).

{\thm \label{asym-beta} Under regularity conditions (a)---(d) in the
Appendix, for the profile least squares estimator of $\bbeta$
defined in (\ref{est}),
\begin{equation}\label{norm-beta}
\sqrt{n}(\hat{\bbeta}-\bbeta)\stackrel{\mathcal{L}}\longrightarrow
N(0,\frac{I}{I-1}\sigma^2\Sigma^{-1}),
\end{equation}
where $\Sigma = \E\Big[\{X-\E(X|Z)\}\{X-\E(X|Z)\}^{\top}\Big]$. }

Similar to the treatment of least square estimator for linear regression models, and
noting that the degrees of freedom of  (\ref{PLinear_Model}) is approximately $(I-1)/I \cdot n$,
we can estimate the variance of $\hat{\bbeta}$ using sandwich formula based on  (\ref{est}).

\begin{equation}\label{beta_err}
\Var(\hat{\bbeta})= \hat{\sigma}^2 \Big\{\sum\limits_{j=1}^J\sum\limits_{i=1}^{I}\{X_{(j-1)I+i}-\frac{1}{I}\sum\limits_{i=1}^{I}X_{(j-1)I+i}\}^T
\{X_{(j-1)I+i}-\frac{1}{I}\sum\limits_{i=1}^{I}X_{(j-1)I+i}\}\Big\}^{-1},
\end{equation}
where
$$
\hat{\sigma}^2= \frac{I}{I-1}\cdot \frac{1}{n} \sum\limits_{j=1}^J\sum\limits_{i=1}^{I}(Y_{(j-1)I+i} -
X_{(j-1)I+i}\, \hat{\bbeta} - \hat{\alpha}_j)^2.
$$

Furthermore, we can plug $\hat{\beta}$ back into equation (2.2) and obtain an updated nonparametric
estimate of $g(Z)$ based on \[ Y_i^*=Y_i-X_i\hat{\beta}\] using standard nonparametric techniques.
Since $\hat{\beta}$ is a root $n$ consistent estimator of $\beta$,  we expect the
updated nonparametric estimator $\hat{g}(Z)$ will converge to $g(Z)$ at the
optimal nonparametric convergence rate.

\subsection{Extension to multivariate nonparametric $g(Z)$}
\emph{Case I:} The simple method of approximating one-dimensional function $g(Z_i)$ can be
readily extended to the multivariate case when $Z$ consists of one continuous variable and
several categorical variables. Note that without loss of generality, we can express multiple
categorical variables as one $K$-level categorical variable. Hence, a partially linear model
\begin{equation}\label{SPL2}
Y_i=X_i\bbeta+g(Z_{i}^d, Z_{i}^c)+\varepsilon_i, \ i=1,\ldots,n,
\end{equation}
where $Z_i=(Z_{i}^d, Z_{i}^c)$ where $Z_{i}^c \in R^1$ as specified in (\ref{SPL}), $Z_{i}^d$
is a $K$-level categorical variable.

To approximate $g(Z^d, Z^c)$ we first split the data into $K$ subsets given the categorical
values of $Z_{i}^d$, then the $k$th ($0\leq k\leq K$) subset of the data will be further
partitioned into sub-intervals of $I$ data points with adjacent values of $Z^c$. Based on the
partition, model (\ref{SPL2}) can still be written in the form of (\ref{PLinear_Model}).
The profile least squares as shown above can be used to estimate $\bbeta$ and we have the
following corollary.

{\cor \label{asym-beta-case1} Under the model (\ref{SPL2}) and regularity conditions (a)---(e),
for the profile least squares estimator
of $\bbeta$ defined in (\ref{est}),
\begin{equation}\label{norm-beta}
\sqrt{n}(\hat{\bbeta}-\bbeta)\stackrel{\mathcal{L}}\longrightarrow
N(0,\frac{I}{I-1}\sigma^2\Sigma^{-1}),
\end{equation}
where $\Sigma = \E\Big[\{X-\E(X|Z)\}\{X-\E(X|Z)\}^{\top}\Big]$. }

\emph{Case II:}
The simple approximation can also be easily applied to continuous bivariate variable
$Z=(Z_1, Z_2)\in R^2$. The partition will need to be done over the bivariate support of $Z$.
In the extreme case when the two components of $Z=(Z_1,Z_2)$ are independent from each
other, the approximation error based on the partition is of order $o(1/\sqrt{n})$, the same
as the model error. Hence in theory the root-$n$ consistency of $\bbeta$ can be established.
Below we outline a corollary that based on the case when the two components of $Z$
are highly correlated so we only need to partition the support of $Z$ according to one component.
First we assume
\begin{equation}\label{closeness}
\Delta_{si}\equiv Z_{1i}-Z_{2i}\rightarrow 0, \quad i=1,\cdots,
n,
\end{equation}
a similar condition as in \citet*{Jiang&Fan&Fan:2010}

Under the assumption (\ref{closeness}) with $\Delta_{si}=o(1)$, it is sufficient to partition
the observations into subintervals of $I$ data points according to the order of
$Z_{1i},i=1,\ldots,n$. If $g(\cdot)$ satisfies some regular smoothness conditions, given
subinterval $j$, $g(\mathbf{Z}_{(j-1)I+i})$ is approximately equal for $i=1, \cdots, I$,
denoted by $\alpha_j$. Again the model can be represented in the form of
(\ref{PLinear_Model}) and we have another corollary,

{\cor \label{asym-beta-case2} Under the model (\ref{SPL2})  where $Z_{1i}$ and $Z_{2i}$ are
highly correlated and satisfy the condition (\ref{closeness}), and the regularity conditions (a)---(d),
for the profile least squares estimator of $\bbeta$
defined in (\ref{est}),
\begin{equation}\label{norm-beta}
\sqrt{n}(\hat{\bbeta}-\bbeta)\stackrel{\mathcal{L}}\longrightarrow
N(0,\frac{I}{I-1}\sigma^2\Sigma^{-1}),
\end{equation}
where $\Sigma = \E\Big[\{X-\E(X|Z)\}\{X-\E(X|Z)\}^{\top}\Big]$. }

The results of the theorem and corollaries are similar as the results of Fan \etal (2005)
except replacing the unconditional asymptotic covariance matrix of the estimate by the conditional
covariance matrix. The proofs of the theorem and corollaries are deferred to the Appendix.


{\emph Remark 1: } We proposed to approximate $g(Z)$ by simply averaging observations within the local neighborhood.
This method to some extent resembles kernel methods with small bandwidth in nonparametric estimation. However,
our method does not require a kernel density function nor complicated bandwidth selection, so it can be viewed
as a ``poor man's" nonparametric method that is completely model free. Theorem \ref{asym-beta} and the
two corollaries demonstrate that the limiting distribution of $\sqrt{n}(\hat{\bbeta}-\bbeta)$  based
on partially consistent estimation of $g(Z)$ is almost as efficient as the
estimator of $\bbeta$ based on classic method for partially linear model
while the latter requires a consistent estimates of $g(Z)$. Theorem 1 shows that the parametric estimates
based on simply a naive approximation of $g(Z)$ can still obtain optimal root-$n$ consistency. In the
extreme case, the consistent estimate of $\bbeta$ can be obtained when the number of observations per
subinterval, $I$, is as small as 2. One only pays the cost in efficiency by a factor of $I/(I-1)$. This
inflation factor diminishes quickly as $I$ increases.

{\emph Remark 2:} Computationally, our method is easy to compute and does not require additional tuning parameter
 selection. In the simulation section we will show the complex computational procedure of the nonparametric kernel method
 also leads to numerical inefficiency, the ratio between the average estimation errors of our method and the nonparametric kernel
 method is in fact less than $I/(I-1)$. In addition, the effectiveness of various kernel functions only depends on
 the underlying assumption about $g(Z)$. In our method, $g(Z)$ is approximated by non-overlapping partitions
 hence it is more local than the kernel methods, and therefore it is more forgiving to oddities such as jumps,
 singularities and boundary effects of $g(Z)$ . Essentially, we do not cast any structural assumption over $g(Z)$
 so it can be more readily extended to deal with multivariate random vectors including ordered and unordered
 categorical data and allow interactions among the components.

{\emph Remark 3:} As the dimension of the continuous components of $Z$ increases, similar as the
discussion of \cite{Fan&Huang:2001} about ordering multivariate vector, $Z$ can be ordered
according to the first principle component of $Z$ or certain covariate. In practice,
as shown by \cite{Cheng&Wu:2013}, the high dimensional continuous random vector $Z$
can often be represented by a low dimensional manifold. Hence we can expect that
for many cases, once $Z$ is expressed in a low dimensional manifold without losing much information,
the partition of $Z$ can be done within the manifold effectively and our results should still apply.
Nevertheless, further investigation is needed to ascertain the conditions needed for the generalization
of our method.

\section{Statistical inference for partially linear regression model}\label{test}

\subsection{Statistical inference for parametric component}

In this section, we investigate statistical inference problem with respect to the estimator of
$\bbeta$. In particular, we consider the following testing problem for $\bbeta$
\begin{equation}
H_0^{1}:A\bbeta=0, \quad  \mbox{vs} \quad   H_1^{1}: A\bbeta\neq0
\end{equation}
where $A$ is a $k\times p$ matrix.  A profile likelihood ratio  or profile least square ratio  test statistic (Fan and Huang, 2005) will be defined
  and we will investigate whether this test statistic is almost efficient and has an easy-to-work limiting distribution

Let $\hat{\bbeta}_0$ be the estimators of $\bbeta$ and
$\hat{\balpha}_{n0}$ be the estimators of $\balpha_n$ in
(\ref{PLinear_Model}) under the null hypothesis $H_0^1$. The
residual sum of squares (RSS) under the null hypothesis is
$\mbox{RSS}_0=n^{-1}\sum\limits_{j=1}^J\sum\limits_{i=1}^I
\hat{\varepsilon}_{(j-1)I+i,0}^2$, where
$\hat{\varepsilon}_{(j-1)I+i,0}=Y_{(j-1)I+i}-\hat{\alpha}_{j0}
-X_{(j-1)I+i}\hat{\bbeta}_0$. Similarly, let $\hat{\bbeta}_1$ and
$\hat{\balpha}_{n1}$ be the estimators of $\bbeta$ and
$\balpha_n$ in (\ref{PLinear_Model}) under the alternative
hypothesis. The RSS under $H_1^1$ is
$\mbox{RSS}_1=n^{-1}\sum\limits_{j=1}^J\sum\limits_{i=1}^I
\hat{\varepsilon}_{(j-1)I+i,1}^2$, where
$\hat{\varepsilon}_{(j-1)I+i,1}=Y_{(j-1)I+i}-\hat{\alpha}_{j1}
-X_{(j-1)I+i}\hat{\bbeta}_{1}$. Following \citet*{Fan&Huang:2005},
we define a profile least squares based test statistic
\begin{equation}\label{pt}
T^1_n = (RSS_0-RSS_1)/RSS_1.
\end{equation}

Then under the regularity conditions and the null hypothesis, we have
the following theorem for the asymptotic distribution of
$T_n^1$.

{\thm \label{asym-T1} Under regularity conditions (a)---(e) in the Appendix,
and given the profile least squares estimator $\hat{\bbeta}_0$ and
$\hat{\bbeta}_1$ defined above,
\begin{equation}
\frac{I-1}{I} \cdot n T^1_n \stackrel{\mathcal{L}}{\longrightarrow} \chi^2_k \quad
\mbox{as} \quad n \to \infty.
\end{equation}
}
For linear regression with normal errors, the same test statistic was shown to
have a Chi-square distribution \citep {Fan&Huang:2005}. The results in Theorem
2 demonstrates that classic hypothesis testing results can still be applied to
the parametric component in (\ref{SPL}) with partially consistent nonparametric
component estimators. The constant $I/(I-1)$ is the price to be paid for
introducing high dimensional nuisance parameters in the model.

In practice, for finite sample size, we can use the following bootstrap procedure to calculate the $p$-value of
the proposed testing statistic $T_n^1$ under null hypothesis.
{\noindent\bf Bootstrap algorithm for $T_n^1$}
\begin{enumerate}[1.]
\item Generate the residuals $\{\hat{\varepsilon}_{(j-1)I+i}^*,j=1,\cdots, J; i=1,\cdots, I\}$
by uniformly resampling from $\{\hat{\varepsilon}_{(j-1)I+i,0}\}$,
then centralize  $\{\hat{\varepsilon}_{(j-1)I+i}^*\}$ to have mean zero.
\item Define the bootstrap sample $Y_{(j-1)I+i}^*=X_{(j-1)I+i}\hat{\bbeta}_0+\hat{\alpha}_{j,0}
    +\hat{\varepsilon}_{(j-1)I+i}^*.$
\item Calculate the bootstrap test statistics $T_n^{1*}$ based on the bootstrap sample
\\$\Big\{(Y_{(j-1)I+i}^*, X_{(j-1)I+i},Z_{(j-1)I+i}), j=1,\cdots,J; \ i=1,\cdots,I\Big\}$.
\item Repeat steps 1-3 to obtain $N$ replicates of bootstrap samples and compute $T_n^{1*,b}$ for
each sample $b=1, \ldots, N$. The $p$-value of the test can be calculated based on the relative
frequency of the events $\{T_n^{1*,b}\geq T_n^1\}$.
\end{enumerate}

\subsection{Statistical inference for nonparametric component when categorical data are involved}

In Corollary 1, we established the root-$n$ consistency of the parametric component $\bbeta$
when the nonparametric component is of the form $Z_i=(Z_{i}^d, Z_{i}^c)$ where $Z_{i}^d$
is a $N$-level categorical variable and $Z_{i}^c$ is a continuous variable in $R^1$.

Given the almost efficient estimate $\hat{\bbeta}$, we have
$Y_i^\ast=Y_i-X_i\hat{\bbeta}=g(Z_{i}^d, Z_{i}^c)+\varepsilon_i^\ast$.
The nonparametric function $g(Z_i)=g(Z_{i}^d, Z_{i}^c)$ can be expressed in terms
of a series of univariate functions conditioning on the values of $Z_{i}^d$,
$g(Z_{i}^c|Z_{i}^d=k), k=1, \ldots, N$. Each of these univariate functions
can be estimated using kernel method based on the split data with corresponding
 $Z_{i}^c$ values. Those estimates can be defined as $\hat{g}(Z_{i}^c|Z_{i}^d=k)$.
 Naturally one likes to test the equivalence of these univariate functions.

Motivated by the real example in Section 5, we consider the following testing problem when $N=2$:
\begin{equation}
\begin{split}
&H_0^{2}: g(Z_{i}^c|Z_{i}^d=0)=g(Z_{i}^c|Z_{i}^d=1) \mbox{\,\, almost everywhere},   \\
& H_1^{2}:g(Z_{i}^c|Z_{i}^d=0) \ne g(Z_{i}^c|Z_{i}^d=1)\mbox{\,\, on a set with
positive measure}.
\end{split}
\end{equation}
The above testing problem resembles a two-population nonparametric testing problem.
For such a testing problem, \citet*{Rachine&Hart&Qi:2006} suggest a quadratic distance
testing statistic. However, the quadratic distance statistics are not sensitive to
the local changes. Based on $L_\infty$ norm and the idea from \citet*{FanZhang:2000}, we
suggest the following statistic in the context of partially linear model.
\begin{equation}\label{npt1}
T_n^2 = (-2\log h)^{1/2}\left[\sup\limits_{Z^c}\frac{
|\hat{g}(Z^c|Z^d=1)-\hat{g}(Z^c|Z^d=0)|}{\sqrt{\widehat{\Var}
\{\hat{g}(Z^c|Z^d=1)-\hat{g}(Z^c|Z^d=0)\}}}-d_n\right],
\end{equation}
where $h$ is the chosen bandwidth parameter when estimating $g(Z^c|Z^d)$ and
$$d_n=(-2\log h)^{1/2}+\frac{1}{(-2\log h)^{1/2}}
\log\Big\{\frac{\int K'^2(t)\,dt}{4\pi\int K^2(t)\,dt}\Big\}, $$
with  $K(\cdot)$ is a kernel function satisfying $\int K(t)\,dt=1$ and $\int
t^2K(t)\,dt>0$.

Notice that $\hat{g}(Z^c|Z^d=1)$ and $\hat{g}(Z^c|Z^d=0)$ are estimated by different samples,
hence $\hat{g}(Z^c|Z^d=1)$ and $\hat{g}(Z^c|Z^d=0)$ can be assumed independent. So
$${\Var}\{\hat{g}(Z^c|Z^d=1)-\hat{g}(Z^c|Z^d=0)\}=\Var\{\hat{g}(Z^c|Z^d=0)\}+\Var\{\hat{g}(Z^c|Z^d=1)\},$$
where $\Var\{\hat{g}(Z^c|Z^d=0)\}$ and $\Var\{\hat{g}(Z^c|Z^d=1)\}$, which can be estimated
using standard nonparametric procedures.

Given the level of the test, when $T_n^2$ is greater than the critical value, $H_0^2$
can be rejected.  In general, critical values can be determined by the asymptotical
distribution of test statistic under the null hypothesis. However, for this kind of nonparametric
testing problem the test statistic tends to converge to its asymptotic distribution very slowly
(\citet{Rachine&Hart&Qi:2006}.) The best way to approximate the null hypothesis distribution
for the above testing statistic is by bootstrapping. Following the idea of
\citet*{Rachine&Hart&Qi:2006}, we suggest a simple bootstrap procedure to approximate the
null hypothesis distribution of $T_n^2$.

{\noindent\bf Bootstrap algorithm for $T_n^2$:}
\begin{enumerate}[1.]
\item Randomly select $Z_{i}^{d*}$ from $\{Z_{i}^d, i=1,\cdots,n\}$ with replacement,
and call $\{Y_i, X_i, Z_{i}^{d*}, Z_{i2}\}$ the bootstrap sample.
\item Use the bootstrap sample to compute the bootstrap statistic $T_n^{2*}$, which is
the same as $T_n^{2}$  except that $Z_{i1}$ is replaced by $Z_{i1}^*$ values.
\item Repeat steps 1 and 2 to obtain $N$ replicates of bootstrap samples and
$T_n^{2*,b}, b=1, \ldots, N$. The $p$-values is based on the relative frequency of the
event $\{T_n^{2*,b}\geq T_n^2\}$  in the replications of the bootstrap sampling.
\end{enumerate}

The distribution of $T_n^2$ under $H_0^2$ is asymptotically approximated by the bootstrap
distribution of $T_n^{2*}$. Now let $Q_{1-\alpha}(T_n^{2*})$ be the $(1-\alpha)$th quantile
of the bootstrapped test statistic distribution, the empirical $(1-\alpha)$ confidence band
for $\{\hat{g}(Z^c|Z^d=1)-\hat{g}(Z^c|Z^d=0)\}$ can be constructed as follows,
\begin{equation}
\Big[\{\hat{g}(Z^c|Z^d=1)-\hat{g}(Z^c|Z^d=0)\}-\Delta_{\alpha}(Z^c),
\,\, \hat{g}(Z^c|Z^d=1)-\hat{g}(Z^c|Z^d=0)\}+\Delta_{\alpha}(Z^c)
\Big],
\end{equation}
where
$$
\Delta_{\alpha}(Z_2)=\{d_n+Q_{1-\alpha}(T_n^{2*})(-2\log
h)^{-1/2}\}\sqrt{\widehat{\Var}\{\hat{g}(Z^c|Z^d=1)-\hat{g}(Z^c|Z^d=0)\}}.
$$


\section{Numerical studies}
\setcounter{equation}{0}
We conduct three simulation examples to examine the effectiveness of the
proposed estimation method and testing procedures for the partially
linear regression model. The first example is a simple partial linear
regression model with one dimensional nonparametric component.
In the second example we consider highly correlated bivariate nonparametric components,
while in the third one, the nonparametric components are mixed
with one categorical and one continuous variable.

To assess estimation accuracy of the parametric components, we compute
mean square error, $\mbox{MSE}(\hat{\beta})=\sum\limits_{l=1}^p (\hat{\beta}_l-\beta_l)^2, $ and
the  average estimation errors, $\mbox{ASE}(\hat{\beta})=\sum\limits_{l=1}^p|\hat{\beta}_l-\beta_l|.$
The robust standard deviation estimate (RSD) of $\hat{\bbeta}$ is calculated using $(Q_3-Q_1)/1.349$
where $Q_1$ and $Q_3$ are the 25\% and 75\% percentiles, respectively. The limiting distributions of
the test statistics $T_n^1$ and $T_n^2$ under the null hypothesis will be simulated. The power curve of each
test will be constructed as well.  Varying the sample size and the size of the subintervals, $I$,
the performance of our proposed estimation and inference methods will be examined and
compared with alternative methods.

For comparison purposes, all the simulations examples are also calculated
using available R packages. Package ``gam" is used to fit generalized additive model,
package ``NP" is used to fit nonparametric regression and packge ``locfit" is used for nonparametric
curve fitting. Generalized cross validation method is used to select the optimal bandwidth whenever it
is applicable.

{\exm \label{example-simple} Consider the following simple partially
linear regression model
$$
Y_i = X_i^{\top}{\beta}+g(Z_i)+\varepsilon_i, \ i=1,\ldots,n,
$$
where $\beta=(1,3,0,0,0,0)$ and $g(Z_i)=3\sin (2Z_i)$.
$(X_i,Z_i), i=1,\ldots,n$  are i.i.d. draws from a multivariate
normal distribution with mean zero and the covariance matrix
$$
\left(
             \begin{array}{cccc}
               1.0 & \rho & \cdots & \rho \\
               \rho & 1.0 & \cdots & \rho \\
              \vdots & \vdots & \ddots & \vdots \\
               \rho & \rho & \cdots & 1.0  \\
             \end{array}
           \right).
$$
with $\rho=0.5$. $\varepsilon_i, i=1,\ldots,n$ are i.i.d. and follow the standard
normal distribution. }

For this example, 400 simulated samples are produced to evaluate the performance of the proposed
parameter estimators. The results will be compared with those produced by function \verb gam ~in
R package  ``gam" that fits Generalized Additive Models.

First the theoretical results of Theorem 1 are nicely illustrated in the left graph of Figure 1
by a linear relationship between $\log(MSE)-\log(I/(I-1))$ and logarithm of the sample size
with a slope close to -1.  Moreover, from Table 1, when the size of the subintervals is moderately
large (e.g. $I>=5$), the average estimation errors (ASE) and the estimated standard error of our
proposed method are comparable with the results based on \verb gam ~function in R. When sample
size increases, the proposed method are better. In the extreme case when $I=2$, the ASE decreases
with sample size and it is only about 1.3 times that of function \verb gam. This implies the empirical
model variance of our method is about 1.7 times that of \verb gam ~results. This and similar results
in the other two numerical studies suggest that although in theory our method has an efficiency
loss by a factor of $I/(I-1)$, in practice the kernel based methods also suffer efficiency loss due to
computational complexity that is not captured in theoretical results.

We also carry out $T_n^1$ to test the following null hypothesis:
$$
H_0^1:\beta_3=\beta_4=\cdots=\beta_p=0.
$$
We examine the size and power of $T_n^1$ by producing random samples from a sequence of
alternative hypothesis models indexed by parameter $\delta_1$ as follows:
$$
H_1^1:\beta_3=\delta_1, \hspace{0.5cm}\beta_l=0 \mbox{\hspace{0.2cm} for\hspace{0.2cm}} l\geq4.
$$ $\delta_1$ takes values from the set $(0,1)$. When $\delta_1=0$, the alternative
hypothesis becomes the null hypothesis. The empirical null distribution of $I/(I-1)nT_n^1$
with $I=2$ for different sample sizes are calculated based on 1000 simulated samples and plotted in
the right panel of Figure 1. We can see that, even for small value of $I=2$, the empirical null
distribution gets closer to the asymptotical distribution $\chi_4^2$ (solid line) as sample size
increases. This is consistent with Theorem 2.

To assess the bootstrap procedures proposed in section 3.1, we generate 1000 bootstrap
samples and calculate the $p$-value of the test for each simulated sample. Figure 2
illustrates the behavior of the power functions with respect to different $\delta$
values and $I$ values. Two sample sizes are considered, the left panel $n=100$, and
the right panel $n=200$. Though small value of $I$ increases the variance of the
estimator, the power of the test $T_n^1$ is not compromised. As shown in Figure 2,
the power curves are similar for different values of $I$. The simulation results further
confirm that the profile least squares test statistic $T_n^1$ is a useful tool for
linear testing problem in the partially linear regression model under partial consistency

{\small\ctable[caption={Average Estimation Errors for Simulation Example 1 (estimated standard errors in parentheses)}, label=
simul1, pos=h!]{lrccccccccccc} {} {            
\FL
Method && \mc{4}{Our Method} &&\mc{1}{GAM}  \NN 
\cmidrule{1-1}\cmidrule{3-6}\cmidrule{8-8}
& &I=2 & I=5 & I=10 & I=20 && \ML
\multirow{1}{*}{n=100}  &  & 0.969(0.316) &0.745(0.258) &0.856(0.284) &
1.095(0.371) && 0.723 (0.231)   \ML                
\multirow{1}{*}{n=200}  &  & 0.662(0.227) &0.520(0.185)
&0.479(0.156) & 0.579(0.177)&& 0.501(0.153) \ML
\multirow{1}{*}{n=400}  &  & 0.456(0.137) &0.344(0.108)
&0.333(0.118) & 0.347(0.123) && 0.348(0.121)
\LL }}

\begin{figure}[!h]
\centering
\begin{tabular}{cc}
\includegraphics[totalheight=1.9in, width=2.4in, origin=l]{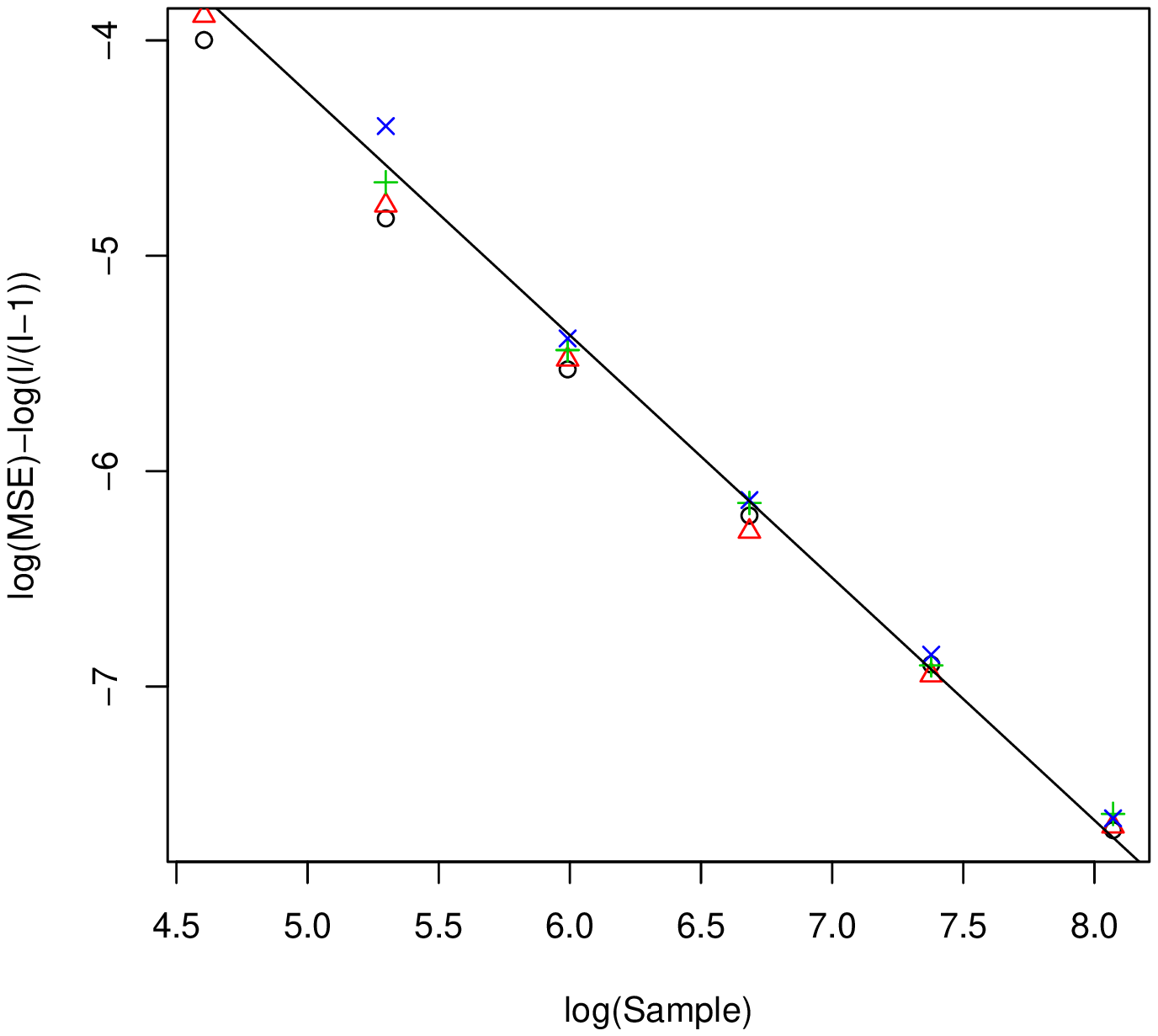}
\includegraphics[totalheight=1.9in, width=2.4in, origin=r]{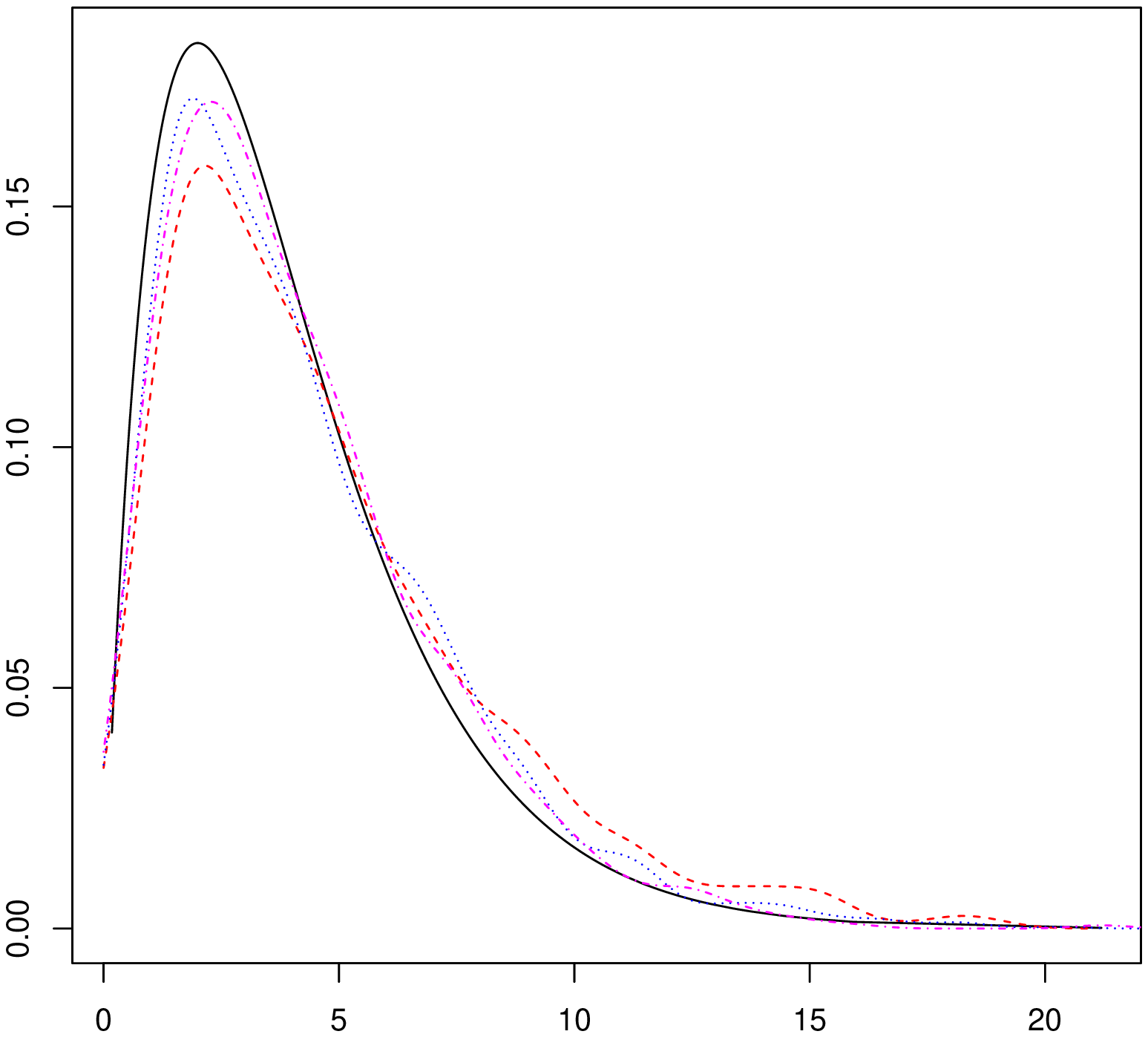} \\
\end{tabular}
\caption{Example 1, Left: Plots of MSEs of $\beta$: I=2 ($\circ$), I=5($\vartriangle$), I=10($+$), I=20($\times$). The slope of the regression line between $\log(\mbox{MSE})-\log(I/(I-1))$ and $\log(\mbox{Sample})$  is -1.12615.
Right: Estimated density of the scaled test Statistic $I/(I-1)nT_n^1$ for
$n=100$ (long-dash), $n=200$ (dot) and $n=400$ (dot-dash) with the $\chi_4^2$ distribution (solid) when $I=2$.     }
\end{figure}

\begin{figure}[!h]
\centering
\begin{tabular}{cc}
\includegraphics[totalheight=1.9in, width=2.4in, origin=l]{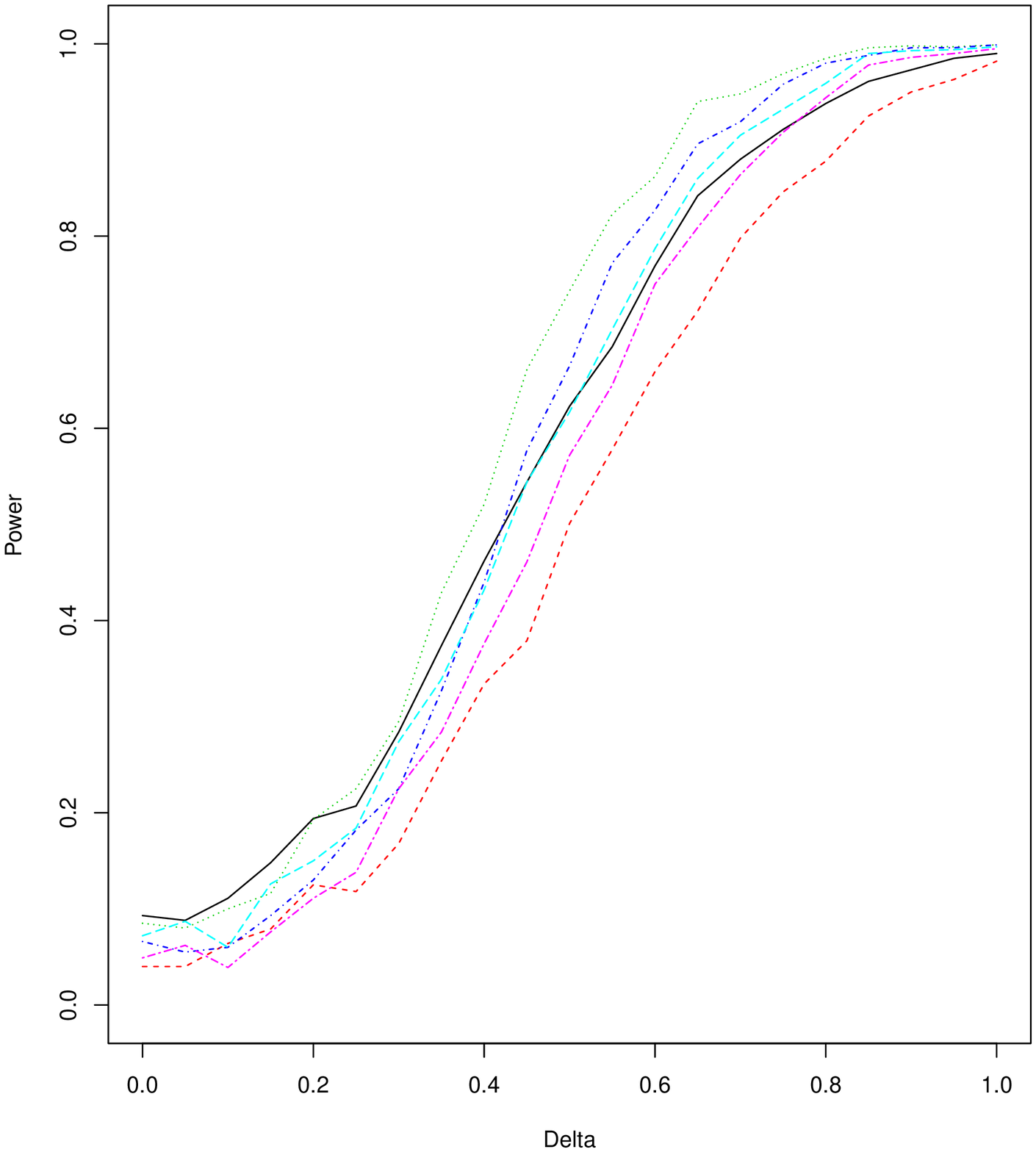}
&
\vspace{-0.25cm}
\includegraphics[totalheight=1.9in, width=2.4in, origin=r]{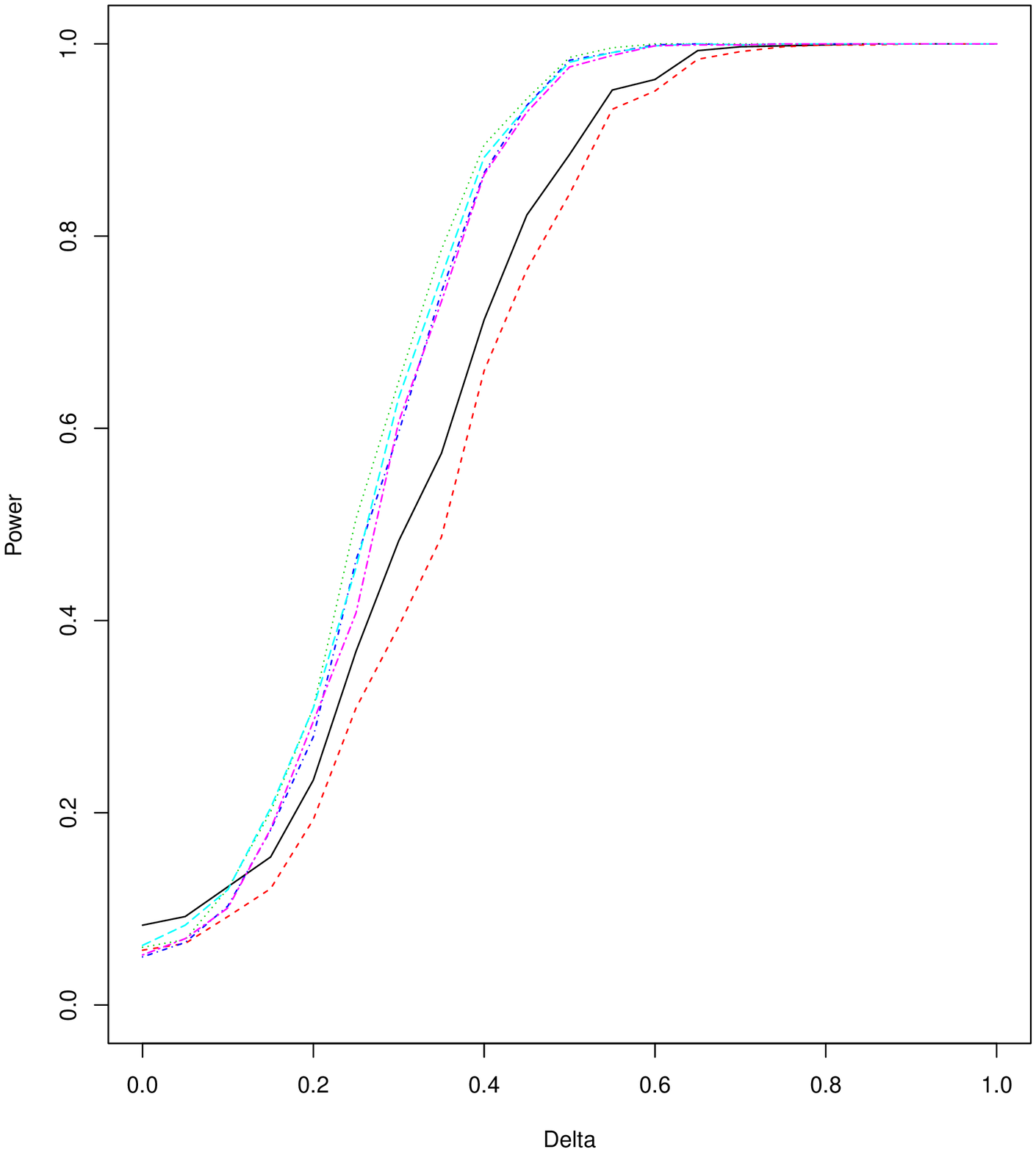} \\
\end{tabular}
\caption{Power of $T_n^1$ for Example 1, Left: $n=100$, Right $n=200$. Solid line(I=2), Dot line (I=5) and Long dash line (I=10) are power curves based on scaled $\chi^2(4)$ distribution. Short dash line (I=2), Dot-Short dash line (I=5), Dot-long dash line (I=10) are  power curves based on the bootstrap algorithm for $T_n^1$. }
\end{figure}

{\exm \label{example-hc} Consider the following generalized additive model,
$$
Y_i =
X_i^{\top}{\beta}+g_1(Z_{1i})+g_2(Z_{2i})+g_3(Z_{3i})+\varepsilon_i,
\ i=1,\ldots, n
$$
where parameter $\beta$ equals to $(1.5,0.3,0,0,0,0)^{\top}$. The functions $g_1, g_2, g_3$ are:
$$
g_1(Z_{1i}) = -5\sin(2Z_{1i}), \hspace{0.5cm}
g_2(Z_{2i})=(Z_{2i})^2-2/3, \hspace{0.5cm}
g_3(Z_{3i})=Z_{3i}.
$$
$X_i$  follows a  multivariate normal distribution with mean vector
zero and the covariance matrix as in Example 1.
The $Z$ are constructed to be highly correlated.
\begin{eqnarray*}
Z_{1} &=& X_{1}+N(0,1)\\
Z_{2} &=& Z_{1}+n^{-1/2}u_1 \\
Z_{3} &=& Z_{1}+n^{-1/2}u_2
\end{eqnarray*}
where $n$ is
the sample size and $u_s$ ($s=1,2$) are $N(0,1)$
variables independent of the covariates. The correlation of $Z$
therefore goes up with sample size. Finally the error term $\varepsilon_i \sim N(0,1)$. }

{\small\ctable[caption={The Average Estimation Errors for Example 2 (estimated standard errors in parentheses)}, label=
simul1, pos=h!]{lrccccccccccc} {} {            
\FL
Method && \mc{4}{Our Method} &&\mc{1}{GAM}  \NN 
\cmidrule{1-1}\cmidrule{3-6}\cmidrule{8-8}
& &I=2 & I=5 & I=10 & I=20 && \ML
\multirow{1}{*}{n=100}  &  &1.375(0.569) &1.305(0.482) &1.636(0.580) &
2.463(0.865) && 1.134 (0.454)   \ML                
\multirow{1}{*}{n=200}  &  & 0.741(261) &0.738(0.257)
&0.876(0.317) & 1.093(0.401)&& 0.791(0.291) \ML
\multirow{1}{*}{n=400}  &  & 0.519(0.184) &0.432(0.161)
&0.473(0.165) & 0.596(0.221) && 0.562(0.213)
\LL }}

\begin{figure}[!h]
\centering
\begin{tabular}{cc}
\includegraphics[totalheight=1.9in, width=2.4in, origin=l]{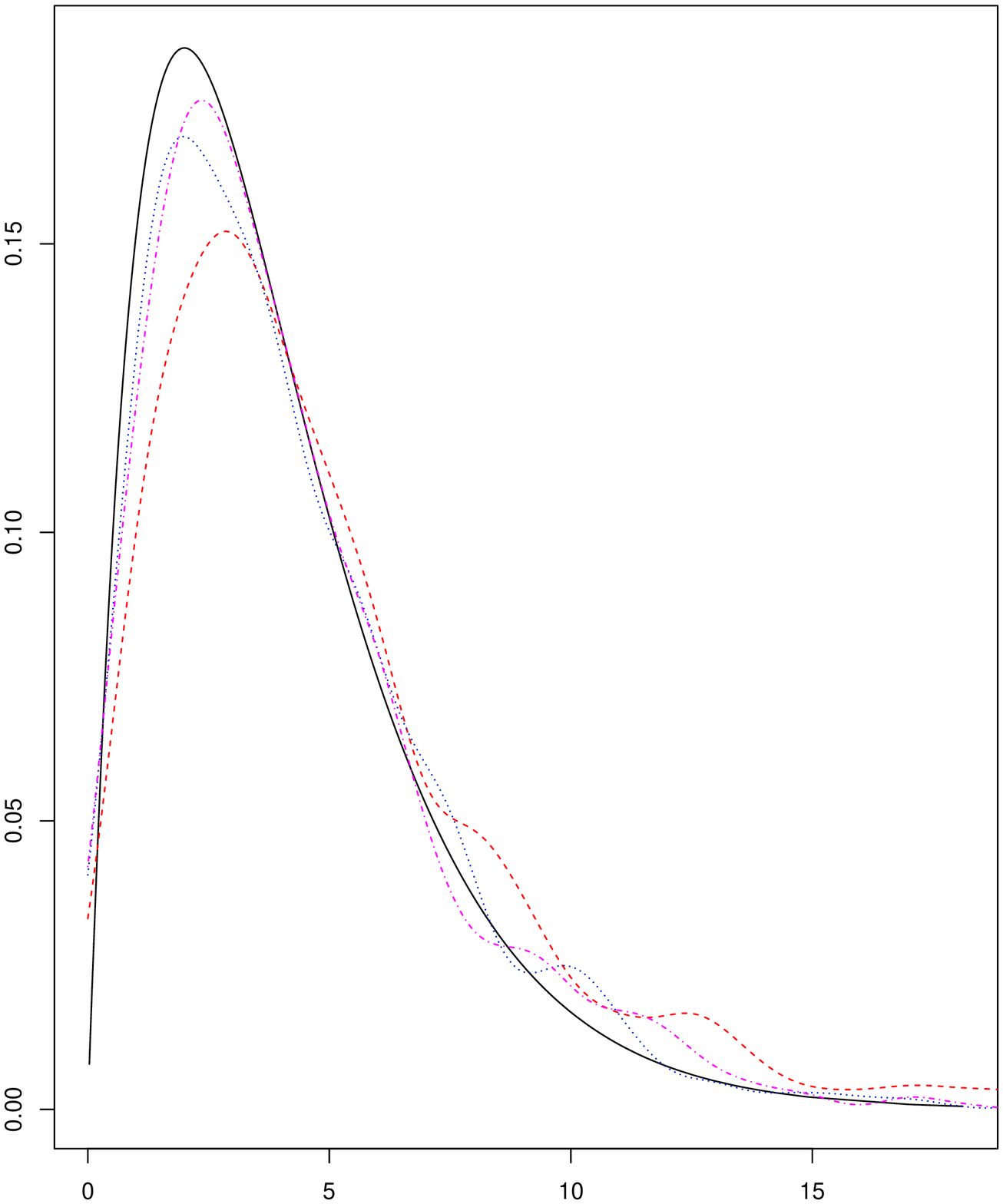}
&
\vspace{-0.25cm}
\includegraphics[totalheight=1.9in, width=2.4in, origin=r]{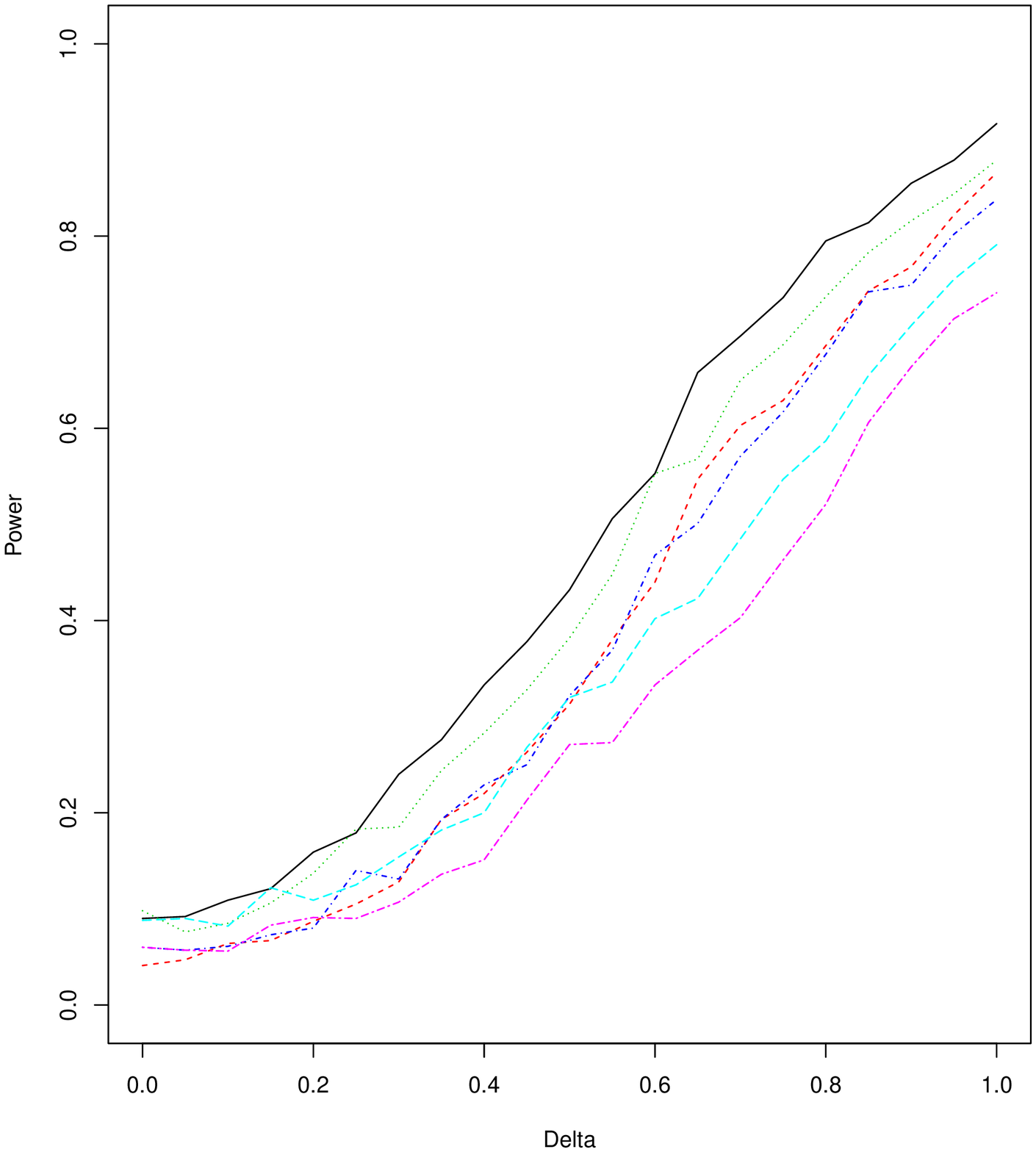} \\
\end{tabular}
\caption{Example 2, Left:  Estimated density of the Test Statistics $I/(I-1)nT_n^1$ for
$n=100$ (long-dash)
, $n=200$ (dot) and $n=400$ (dot-dash) with the $\chi_4^2$ distribution (solid) when $I=2$.     Right: Power of $T_n^1$,  Solid line(I=2), Dot line (I=5) and Long dash line (I=10) are power curves based on $\chi^2_4$ distribution. Short dash line (I=2), Dot-Short dash line (I=5), Dot-long dash line (I=10) are  power curves based on the bootstrap algorithm for $T_n^1$. }
\end{figure}

As in Example 1, 400 simulation examples are used to evaluate the performance of the
proposed estimating method. One thousand simulation examples and the same number of bootstrap samples
are used to study the properties of $T_n^1$ for the same testing problem investigated in Example 1.
As indicated by Table 2, as sample size increases, our proposed method outperforms the \verb gam  ~package
even when $I=2$.  In general, we can see that the proposed method is not sensitive to the choice of $I$
as long as it is not chosen to be too large a value relative to the sample size. Given a fixed sample
size, larger $I$ will yield smaller number of subintervals and lead to coarser approximation of the
nonparametric function. The empirical null distribution of $(I-1)/I n T_n^1$ in comparison with $\chi^2_4$ is
shown in Figure~3. It can be seen that as in Example 1, the empirical null distribution is
a reasonable approximation of the asymptotical null distribution $\chi_4^2$. This is true for various
values of $I$. It is also consistent with the result of Theorem 2.

Compared with the results in Example 1, additional nonparametric component increases the
estimation variability for our proposed method and the method of GAM. ASE and standard
errors are larger in Example 2. It also reduces the power of $T_n^1$ for the same testing
problem as shown in the right graph of  Figure 2. However, our proposed method is more
robust to the high correlation situation as it is able to produce more efficient
results than \verb gam ~when sample size increases.

{\exm \label{example-cate} The model is
$$
Y_i = X_i^{\top}\beta+g(Z_{i}^d,Z_{2}^c)+\varepsilon_i, \ i=1,\ldots,n.
$$
where
$$
g(Z_{i}^d,Z_{i}^c)=(Z_{i}^c)^2+2Z_{i}^c+0.25Z_{i}^de^{-16Z_{i}^{c2}}.
$$
and the true parameter $\beta$ is a $6 \times 1$ vector and equals to
$(3.5,1.3,0,\cdots,0)^{\top}$. $X_i, i=1,\ldots, n$  are independently
generated from Bernoulli distribution with equal probability being 0 or 1.
The categorical variable $Z_{i}^d$ is a Bernoulli variable independent of
$X_i$ with $P(Z_{i}^d=1)=0.7$. The variable $Z_{i}^c$ is continuous and sampled
from a uniform distribution on $[-1,1]$ and independent of $X_i$ and $Z_{i}^d$.
The error term $\varepsilon\sim N(0,0.2^2)$. }

For comparison purpose, we use R package \verb np ~to estimate the bivariate function
$g(Z_{i}^d, Z_{i}^c)$ nonparametrically. In addition, we also use package \verb gam ~to
estimate a ``pseudo" model with an additive nonparametric structure specified as below,
{\label{example-pseudo}
$$
g(Z_{i}^d, Z_{i}^c)= \delta Z_{i}^d+ g(Z_{i}^c)+\varepsilon_i, \ i=1,\ldots,n.
$$}
The true nonparametric components are plotted in the left panel of Figure 4. We can see
that the ``pseudo'' model misspecifies the nonparametric components. It will be interesting to
compare the performance of the proposed method, generalized additive model and nonparametric method
in terms of estimation of the parametric parameter $\bbeta$.

Again, we produced 400 samples for numerical comparison. Table 3 presents the ASE and estimates of
$\bbeta$ under three different methods. The \verb np ~method tries to estimate $\bbeta$ and the bivariate
function $g(Z_1, Z_2)$ simultaneously which involves iterative algorithm and complicated tuning
parameter selections. Hence we expect the numerical performance will be compromised to some extent.
As the other two simulation studies suggested, our method in general produces slightly bigger
ASE than the \verb np ~method but in a factor less than $I/(I-1)$. On the other hand our method
produces more precise estimates of $\bbeta$ than the nonparametric approach. It is
interesting that the GAM approach outperforms the nonparametric approach even under
the wrong model specification. In the left panel of Figure 4, we can see that the difference
between curves $g(Z_{i}^c, Z_{i}^d=0)$ and $g(Z_{i}^c, Z_{i}^d=1)$ is small relative to the
noise hence the more parsimonious specification of the nonparametric part to some extent
improves the parametric estimation. However, under the GAM model wrong inference regard
the nonparametric components will be made.

In Table 4 we compare the empirical standard deviation of $\hat{\bbeta}$ (SD$_m$) with the one
calculated under proposed sandwich formula (\ref{beta_err})(SD). It is obvious that our proposed formula
provide a consistent estimate of the standard deviation of the estimate $\hat{\bbeta}$.

{\tiny\ctable[caption={Fitting Results of ASE and Estimation of $\bbeta$ for Example 3 based on the proposed method, NP and GAM
(estimated standard errors in parentheses)}, label=
simul3, pos=h!]{lrcccccccccccc} {} {          
\FL
Method & \  & \mc{4}{Our Method} &&\mc{1}{NP}  && \mc{1}{GAM} \NN 
\cmidrule{1-1}\cmidrule{3-6}\cmidrule{8-8}\cmidrule{10-10}
\mc{1}{n} & &I=2 & I=5 & I=10 & I=20 & & &   \ML
\mc{1}{100}  &  ASE  & 0.302 (0.104) & 0.298(0.091)  &0.367(0.118)  & 0.505(0.165) && 0.254 (0.091) && 0.217 (0.078)    \ML
\      & $\beta_1$              & 3.504(0.067)  & 3.502(0.063)  &3.506(0.076)  & 3.498 (0.112)&& 3.464 (0.058) && 3.499(0.047)   \ML
\ & $\beta_2$                    & 1.305 (0.067) & 1.294(0.065)  &1.302(0.086)  & 1.310(0.100) && 1.291(0.055)  && 1.302 (0.047)  \ML
\mc{1}{200}  &  ASE  & 0.197(0.064)  & 0.163(0.052)  &0.187(0.059)  & 0.242(0.072) && 0.153 (0.052) && 0.149(0.048)   \ML
\ & $\beta_1$                   &  3.502(0.045) & 3.497 (0.033) & 3.498(0.042) & 3.504 (0.055)&& 3.486 (0.032) &&  3.499(0.030)   \ML
\ & $\beta_2$                    & 1.300(0.041)  & 1.299(0.035)  & 1.303(0.039) & 1.297 (0.054)&& 1.293(0.031)  && 1.299(0.032) \ML
\mc{1}{400}  &  ASE  &0.138 (0.041)) & 0.113(0.037)  &0.105(0.035)  & 0.121(0.042) && 0.102(0.032)  && 0.108 (0.037)   \ML
\ & $\beta_1$                   & 3.500(0.029)  & 3.499(0.024)  & 3.500(0.022) & 3.501(0.027) && 3.492(0.022)  && 3.497 (0.021)      \ML
\ & $\beta_2$                   &  1.303(0.031) & 1.298(0.022)  & 1.300(0.023) & 1.300(0.024) && 1.300 (0.024) && 1.300 (0.023)
\LL }}

{\tiny \ctable[caption={Standard Deviations of Estimates $\hat{\beta}_1$ and $\hat{\beta}_2$ in Example 3  (estimated standard errors in parentheses)}, label=
simul3err, pos=h!]{lrccccccccccc} {} {          
\FL
\ & \  & \mc{2}{I=2} &&\mc{2}{I=5}  && \mc{2}{I=10} && \mc{2}{I=20} \NN 
\cmidrule{3-4}\cmidrule{6-7}\cmidrule{9-10} \cmidrule{12-13}
 n  & \ &    SD & SD$_{m}$  (SD$_{mad}$) &  & SD & SD$_{m}$ (SD$_{mad}$) & & SD & SD$_{m}$(SD$_{mad}$) & & SD &  SD$_{m}$(SD$_{mad}$) \ML
\multirow{1}{*}{100}  &   $\beta_1$  & 0.067 & 0.061(0.0088) & &0.063 & 0.057(0.0067) & & 0.076& 0.071(0.0073)& & 0.112 &0.096(0.0105)       \ML
\ & $\beta_2$   &0.067&0.062(0.0079) & &0.065&0.058(0.0063) & & 0.086 & 0.071 (0.0079)& & 0.100& 0.096(0.0112)           \ML
\multirow{1}{*}{200}  &  $\beta_1$ & 0.045 & 0.041(0.0037) & &0.033 &0.035(0.0021) & & 0.042 &0.037(0.0024) & & 0.055 &0.048(0.0046)           \ML
\ & $\beta_2$      & 0.041 &0.041(0.0038)  & & 0.035 & 0.034(0.0022)& & 0.039 &0.037(0.0024) & & 0.054 & 0.048(0.0044)              \ML
\multirow{1}{*}{400}  &   $\beta_1$ & 0.029 & 0.028(0.0018) & & 0.024 &0.023(0.0010) & & 0.022 &0.023(0.0009) & & 0.027 & 0.025(0.0010)         \ML
\ & $\beta_2$ & 0.031 &0.028(0.0018) & & 0.022 & 0.023(0.0011) & & 0.023 & 0.023(0.0009) & & 0.024 &0.025(0.0010)
\LL }}

Next we test the equivalence of the two nonparametric components associated with $Z^d=0,1$,
\begin{equation*}
\begin{split}
&H_0^2: g(Z^c, Z^d=0)=g(Z^c, Z^d=1),\\
&H_1^2: g(Z^c, Z^d=0) \ne g(Z^c, Z^d=1)
\end{split}
\end{equation*}
In this simulation example, $g(Z_{i}^c, Z_{i}^d=0)=(Z_{i})^{c2}+2Z_{i}^c$ and
$g(Z_{i}^c, Z_{i}^d=1)=(Z_{i})^{c2}+2Z_{i}^c+\delta\exp(-16(Z_{i}^{c2}))$.
To explore the relationship between effect size and power of our proposed
test statistic $T_n^2$, we let the value of $\delta$ change from 0 to 0.25.

To calculate $T_n^2$, we first get the estimate of $\bbeta$, $\hat{\bbeta}$ using formula (2.5),
then remove it from the model,
$$
Y^\ast_i=g(Z_{i}^c, Z_{i}^d) +\varepsilon^\ast_i, \quad i=1,\ldots,n
$$
where $Y_i^\ast=Y_i-X_i\hat{\bbeta}$ and $\varepsilon_i^\ast=\varepsilon_i+X_i\bbeta-X_i\hat{\bbeta}$.
Next we use R package \verb locfit ~to select a bandwidth $h$ and in fact use $0.8h$ to get slightly
under-smoothed estimates of $\hat{g}(Z_{i}^c, Z_{i}^d=0)$ and $\hat{g}(Z_{i}^c, Z_{i}^d=1)$ and
their variance estimates. The test statistic $T_n^2$ is calculated by plugging these estimators
into formula (\ref{npt1}). The $p$-values associated with $T_n^2$ are calculated using the
bootstrap procedure suggested in Section 3. One thousand bootstrap samples are used to approximate
the null distribution of $T_n^2$. This procedure is repeated 400 times to calculate the
power of the test statistic under the alternative models defined by various
$\delta$ values from 0 to 0.25.

The empirical distribution and bootstrapped distribution of $T_n^2$ under null hypothesis when $\delta=0$ and the bootstrapped null distribution approximation under alternative hypothesis when $\delta=0.083, 0.167, 0.25$  at sample size
$n=200$ and $I=5$ are shown in the middle graph of Figure 4. We can see that the bootstrapped distributions under different alternative models provide
 fairly good approximations to the real null hypothesis distribution of our proposed test statistics. It suggests that the asymptotical null distribution of the proposed test statistics for our two population nonparametric testing problem should be a model free test statistic.
In the right panel of Figure 4, the power curves of $T_n^2$ under various $\delta$ values and different sample sizes are shown.
The estimates of $\bbeta$ have little impact on the power curves and such impact is only through sample size.
As a two-population nonparametric test, it is not too surprising to see that the power of this test is
relatively low for small sample size. But as the sample size doubles, the power function picks up quickly even for small effect size when $\delta=0.1$.

\begin{figure}[!h]
\centering
\begin{tabular}{cc}
\includegraphics[totalheight=1.5in, width=2in, origin=l]{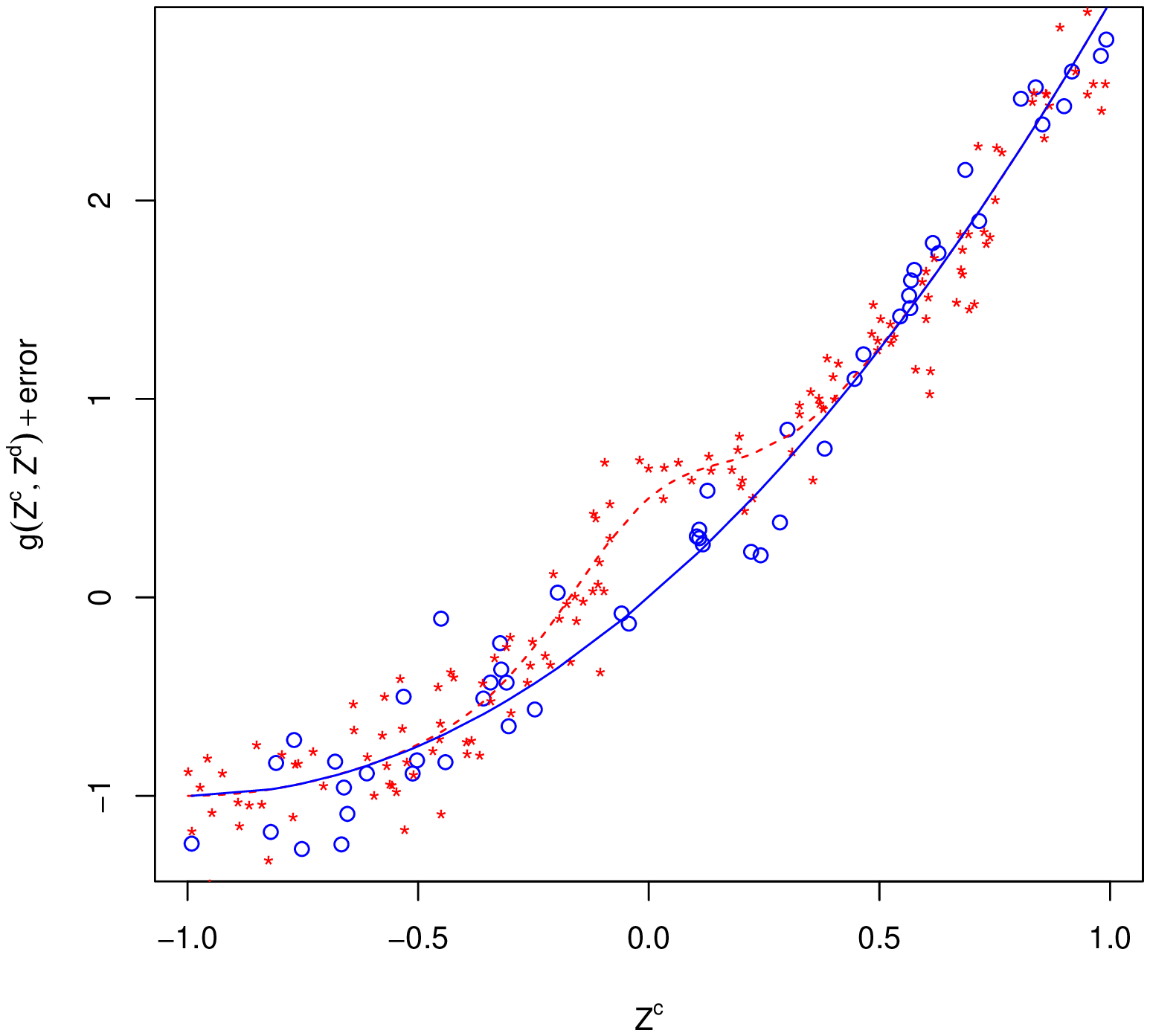}
&
\vspace{-0.25cm}
\includegraphics[totalheight=1.5in, width=2in, origin=l]{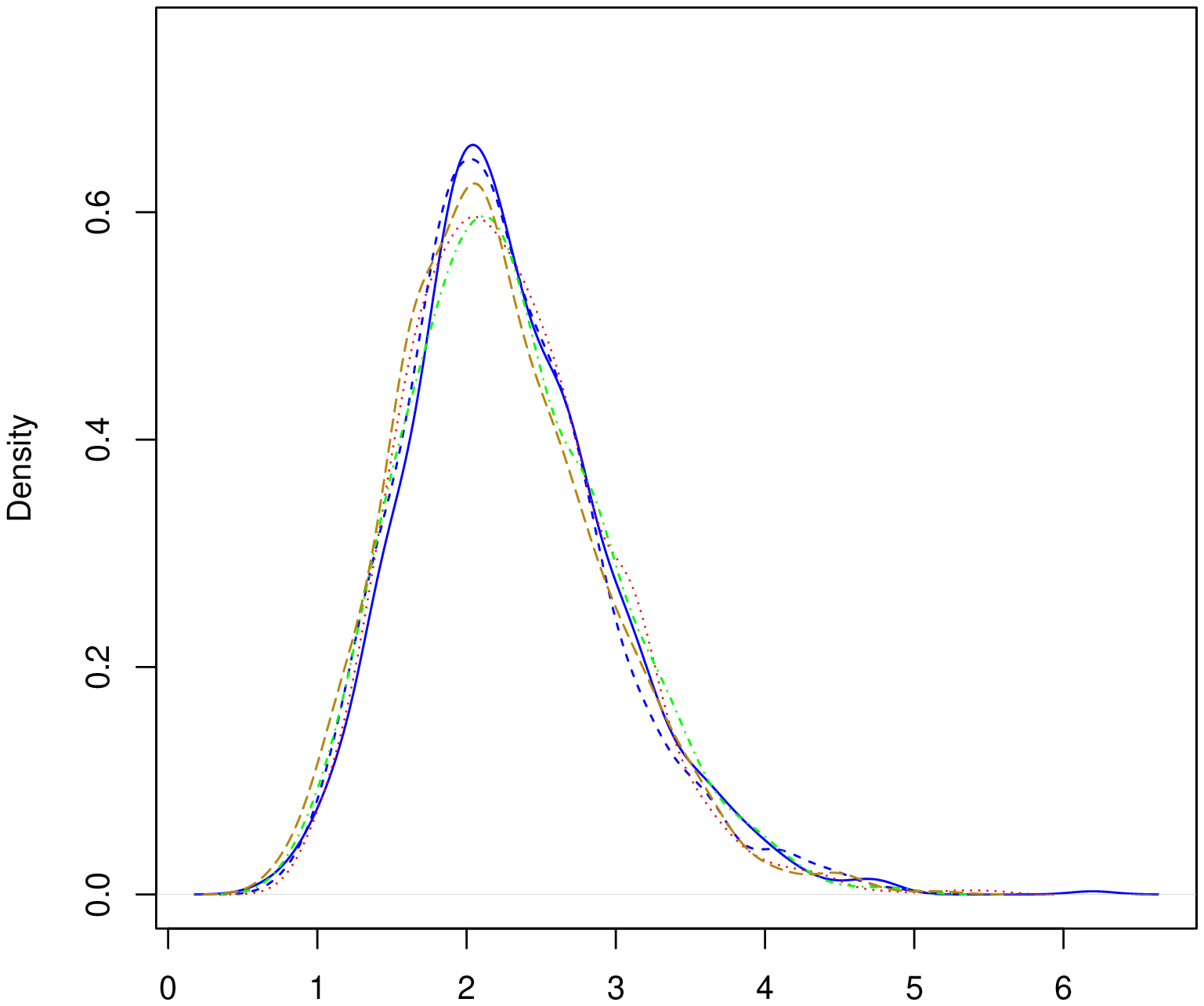}
\vspace{-0.25cm}
\includegraphics[totalheight=1.5in, width=2in, origin=l]{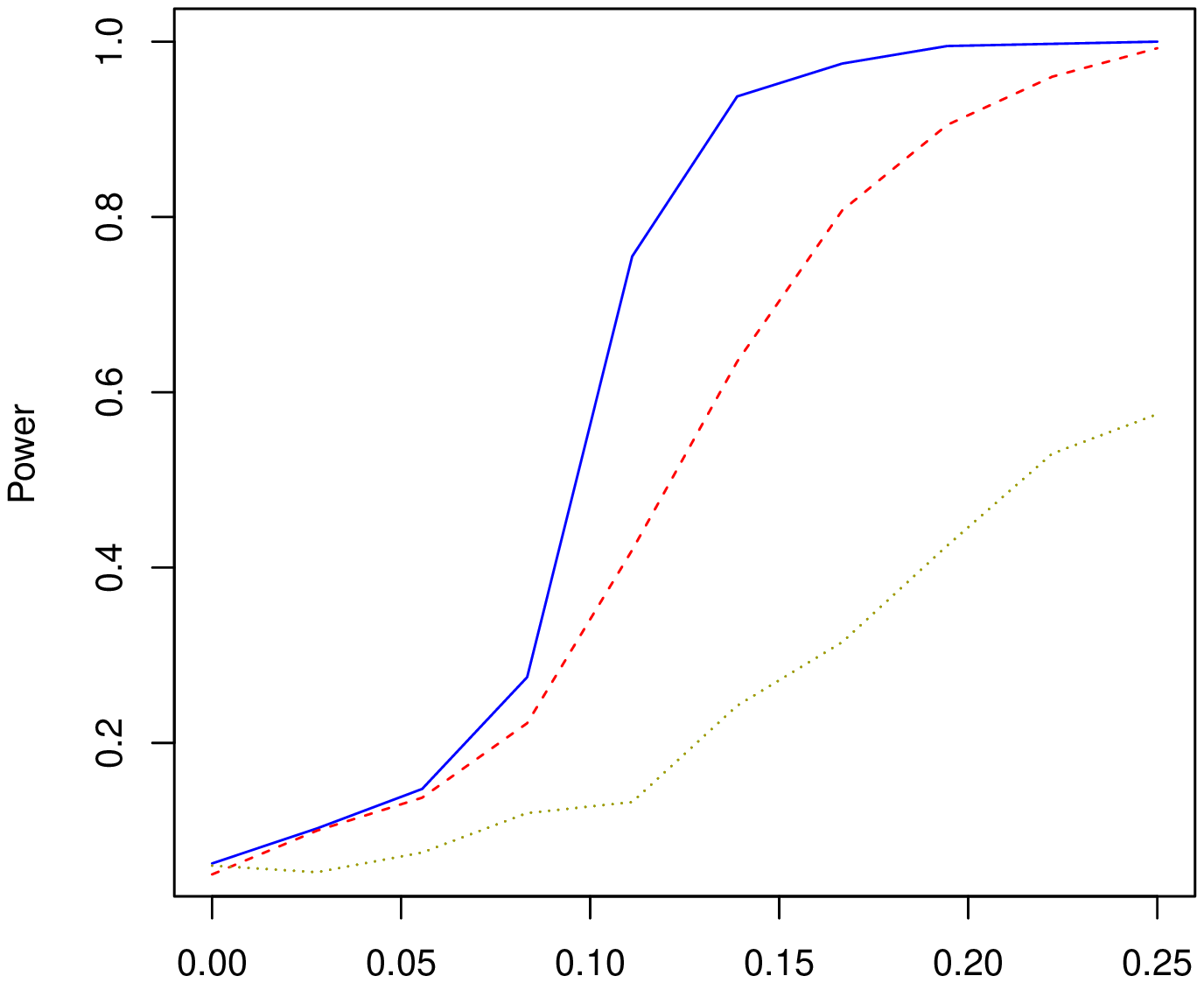}
\\
\end{tabular}
\caption{Example 3, Left: Scatterplot of $g(Z^c,Z^d)+$error vs $Z_2$ overlaid by
solid-blue line: $g(Z^c, Z^d=0)$ and dash-red line: $g(Z^c, Z_d=1)$.
 Middle:  Estimated density of the empirical and bootstrap null distribution
 of $nT_n^2$ for $n=200$ and $I=5$: solid blue line ($\delta=0$) is the empirical null distribution. Dash-blue line ($\delta=0$), dot-red line ($\delta=0.083$), dot-dash green line ($\delta=0.167$)
 and long-dash dark golden red line ($\delta=0.25$) are bootstrapped estimation of null distribution.
 Right: the power function evaluated at $I=5$ and  different $\delta$ values with different sample sizes $n=100$ (dot line), $n=200$ (dash line) and $n=400$ (solid line). }
\end{figure}

\section{Real data application: correlates of birth weight}
\setcounter{equation}{0}
\
Low birth weight is an important biological indicator since it is associated
with both birth defects and infant mortality. A woman's physical condition
and behavior during pregnancy can greatly affect the birth weight of the newborn.
In this section, we apply our proposed methods to a classic example studying the
determinants of birth weights (Hosmer and Lemeshow, 2000). This dataset is part of
a larger study conducted at Bay State Medical Center in Springfield, Massachusetts.
The dataset contains variables (see below) that are believed to be associated with low birth
weight in the obstetrical literatures. The goal of the analysis is to determine
whether these variables are risk factors in the clinical population being served
by Bay State Medical Center.
\begin{itemize}
\item MOTH\_AGE:		Mother's age (years)
\item MOTH\_WT: 		Mother's weight (pounds)
\item Black:		Mother's race being black ('White' is the reference group)
\item Other:        Mother's race being other than black or white
\item SMOKE:		Mother's smoking status (1=Yes, 0=No)
\item PRETERM:		    Any history of premature labor (1=Yes, 0=No)
\item HYPER:			History of hypertension (1=Yes, 0=No)
\item URIN\_IRR:		History of urinary irritation (1=Yes, 0=No)
\item PHYS\_VIS:		Number of physician visits
\item BIRTH\_WT		Birth weight of new born (grams)
\end{itemize}

First we analyze this data set using a linear regression model to estimate the
relationship between various factors and birth weight. Shown in Table 5 (OLS-1 model),
mother's race (Black vs White, Other vs White), history of pregnancy hypertension and history of
urinary irritation have significantly negative impact on birth weights of newborns,
while mother's weight is positively related to birth weight. Perhaps
surprisingly, mother's age is not a significant predictor of baby's birth weight(p-value=0.30).
To check the linearity assumption with respect to the two continuous predictors,
mother's age and weight, standardized residual plot against each of them is
examined. Figure 5a shows that linearity is an adequate assumption for mother's weight
and this relationship is not different between smokers and nonsmokers. But the residual
diagnostics (graph not shown) indicate that the relationship between mother's age and birth
weight is not linear and the relationship could potentially vary by mother's smoking status.

Then we expand the analysis to 1) a linear regression with interaction term between age and
smoking (the OLS-2 model), and 2) a generalized additive model (GAM) that specifies a
nonparametric term with respect to mother's age. Under the OLS-2 model, the baseline
age effect is insignificant. Although the interaction term improves the model fit slightly,
it is deemed insignificant (p-value=0.12). Under the GAM model, the nonparametric term
of age is also tested insignificant (p-value=0.56). The conclusions about the effects of other
variables on birth weights are similar compared to the OLS-1 model.

\ctable[caption = Estimated effects of correlates of birth weight and their standard errors, label=real-data, pos=!h]
{cccccc} {} {
\FL
&                        & OLS-1           & OLS-2          & GAM & PL       \ML
& Intercept              & 3026.9(308.2) & 2741.9(357.0)& 3044.2(309.0)& 2482.0(388.2) \NN
& MOTH\_WT                & 4.6(1.7)      & 4.5(1.7)  & 4.5(1.7)& 5.6(2.0)   \NN
& Black                  & -482.2(146.8) & -431.5(149.7)  &-480.1(147.4)& -295.2(175.2)   \NN
& Other                  & -327.5(112.6) & -302.2(113.3) &-320.1(112.9) & -203.6(132.8)   \NN
& PRETERM                   & -179.5(133.8) & -169.8(133.4) &-166.4(134.2) & -220.0(153.5) \NN
& HYPER                   & -584.4(197.6) & -588.4(196.8) &-582.2(198.1)& -651.7(232.2)\NN
& URIN\_IRR               & -492.3(134.6) & -526.1(135.8)&-508.2(134.9)& -510.2(153.6)\NN
& PHYS\_VIS               & -7.0(45.4)    & -0.7(45.4) &-12.2(45.5)& -14.7(52.8)  \NN
& MOTH\_AGE               & -10.4(9.9)    & 1.8(12.6)    & ---(---)& ---(---)       \NN
& SMOKE                  & -312.5(104.5) & 402.1(468.5)  & -321.3(104.7)& ---(---)   \NN
& MOTH\_AGE $\times$ SMOKE  & ---(---)        & -30.6(19.6) & ---(---)& ---(---)  \ML
& $R^2$                  & 0.251         & 0.261     & 0.255 & 0.391       \LL
}

To model the nonlinear relationship between age and birth weight as well as
its interaction with mother's smoking status, we further fit this data to a partially
linear model with a bivariate nonparametric components, specified as,
\begin{equation}\label{PL}
\begin{array}{rll}
\mbox{BirthWT} &= &\beta_0 +
                \beta_1\mbox{MOTH\_WT} + \beta_2 \mbox{Black} + \beta_3 \mbox{Other}
                + \beta_4\mbox{PRETERM} + \beta_5 \mbox{HYPER} \\
               && + \beta_6 \mbox{URIN\_IRR} + \beta_7 \mbox{PHYS\_VIS}
                + g(\mathrm{MOTH\_AGE},\mathrm{SMOKE}) + \varepsilon.
\end{array}
\end{equation}

We then fit this model using the method proposed in Section 2.3. Since mother's
age is recorded by a series of discrete values from 14 to 36 years, we first partition the
support of $g(\mbox{MOTH\_AGE},\mbox{SMOKE})$ according to mother's smoking status, then
estimate the nonparametric response curve for each group at every distinct age using available
sample points (instead of using fixed cell size). The parameter estimates of the parametric
components with standard errors are given in the last column of Table 5.

Given the parametric components, we re-estimate $g(\mbox{MOTH\_AGE},\mbox{SMOKE})$
using the local polynomial regression methods via \verb locfit. The fitted curves
(after removing the parametric components) are shown in the right panel of Figure 5.
This figure reveals that the response curves between age and birth weight are quite
different for smoking and nonsmoking mothers. We can see that among non-smoking
mothers, age is not particularly associated with birth weight. However, for
smoking mothers, the birth weight decreases quite dramatically as mother's age
increases. The gap is as wide as over 400 grams of birth weight between nonsmoking
and smoking mothers who are 30 years and older. Similar as in the simulation studies,
in the local polynomial regression (\verb locfit), a quadratic term is used
and the optimal bandwidth is chosen via generalized cross validation.

We also conduct the following one-sided nonparametric test to compare
the two response curves between smokers and nonsmokers,

\begin{equation}\hspace{-1cm}
\begin{split}
&H_0^2:g(\mbox{MOTH\_AGE},\mbox{Smoke})= g(\mbox{MOTH\_AGE},\mbox{Nonsmoke}), \mbox{\,\, almost everywhere}\\
&H_1^2:g(\mbox{MOTH\_AGE},\mbox{Smoke})< g(\mbox{MOTH\_AGE},\mbox{Nonsmoke}), \mbox{\,\, on a set with positive measure}.
\end{split}
\end{equation}

Based on (3.16), the test statistic $T_n^2$ for the above test is 3.36,
and the bootstrap p-value is 0.029, suggesting that the response curve of age
among smokers is lower than that of non-smokers. Taking this result and
Figure 5b, we can see that the PL model provides a better specification for
the relationship between mother's age, smoking status and birth weight.

\begin{figure}[!h]\label{conf-curve}
\centering
\includegraphics[totalheight=1.8in, width=2.2in, origin=l]{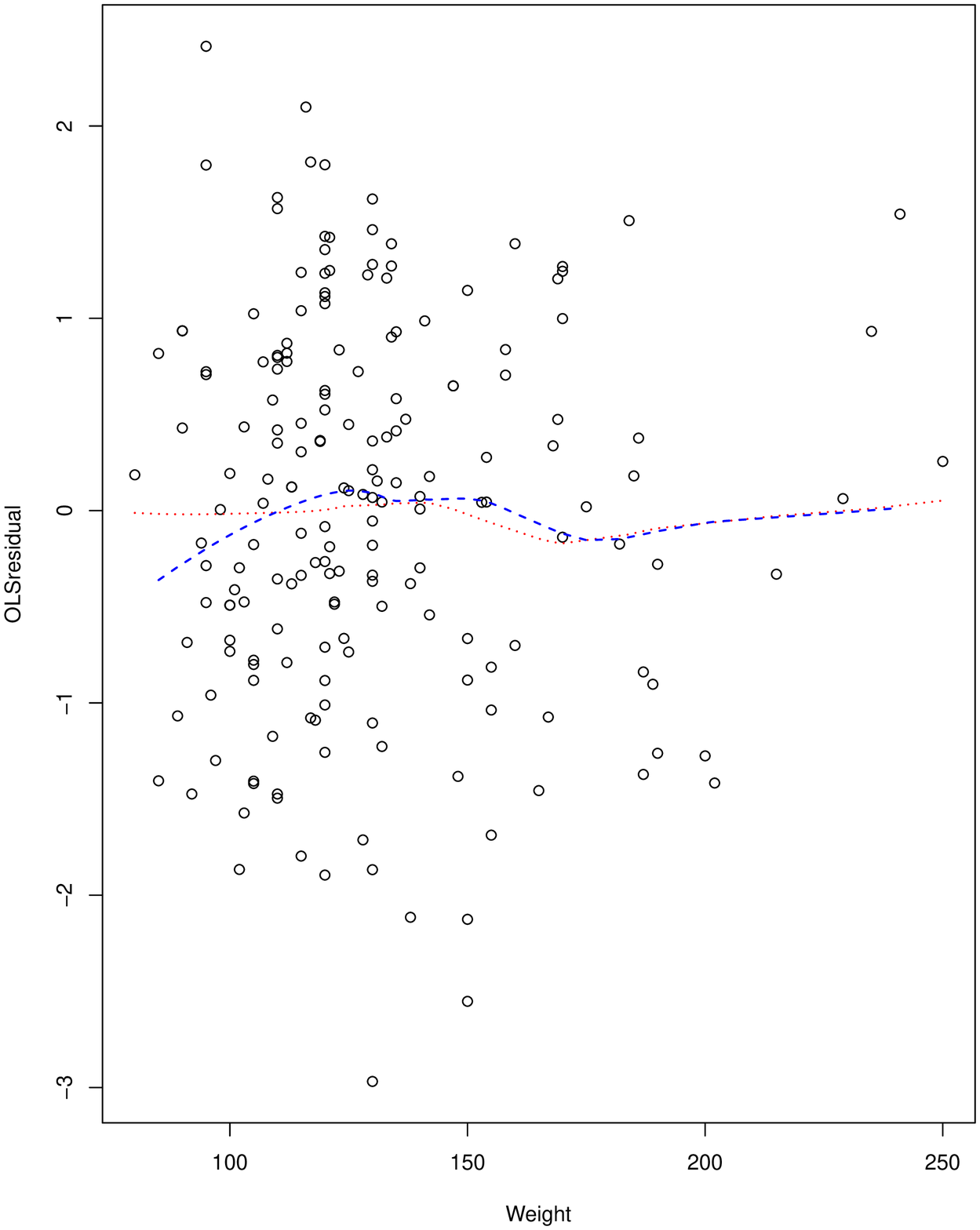}\hspace{-0.4cm}
\includegraphics[totalheight=1.8in, width=2.2in, origin=r]{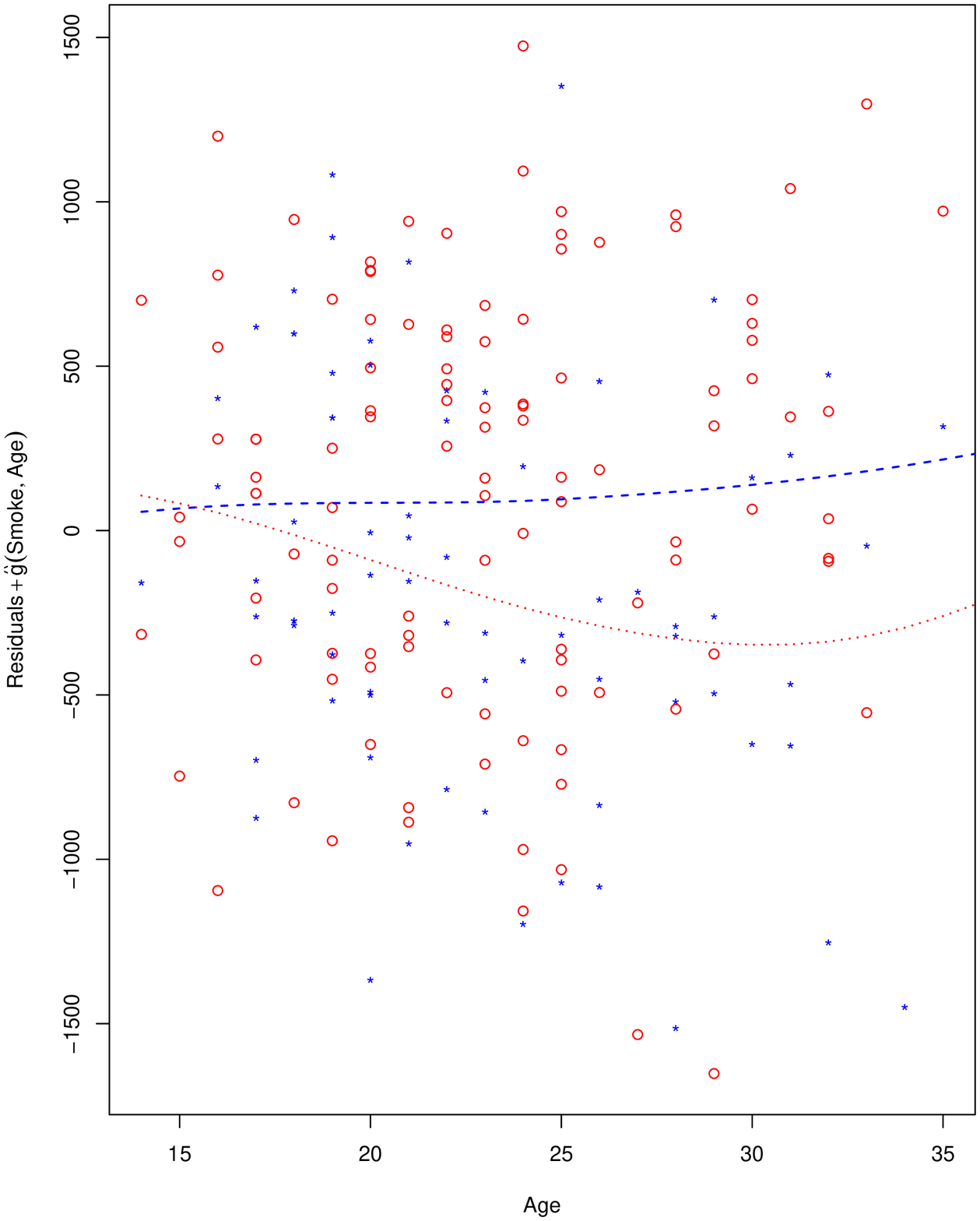}
\caption{\small Correlates of birth weight: The left graph plots the residuals (OLS-2) vs mother's weight. The dotted red and dashed blue lines are the lowess fits for smoking mothers and non-smoking mothers, respectively.  The right graph plots estimated regression function $g(\mbox{Age}, \mbox{Smoke})$ removing the effects of other covariates under the partial consistency PL model. The dotted red and dashed blues are the (lcofit) nonparametric estimates of response curves for smoking and non-smoking mothers, respectively.}
\end{figure}

The estimates of the parameter components of the PL model also exhibit some interesting
changes compared with other models. We can see the racial gap in birth weight narrows.
Controlling other factors, on average babies born to Black mothers are 295 grams lighter than those born to
White mothers. This difference is much smaller compared to the previous models. In addition,
based on the test statistic defined in (3.13), the effect of "Black" now is only marginally
significant (p-value=0.1) and the effect of "Other" becomes insignificant (p-value=0.147).
The effect sizes and significance values of other covariates remain about the same.

\section{Discussion}

In this paper, based on the concept of partial consistency, we proposed a simple estimation
method to partially linear regression model. The nonparametric component of the model is transformed into a set of artificially created nuisance or incidental parameters. Though these nuisance parameters cannot be estimated consistently, the parametric components of the partially linear model can be estimated consistently and almost efficiently under this configuration. As long as the sample size is reasonably large, the number of the nuisance parameters used is not too important. The estimation results have been shown to be fairly stable under various ``coarseness" of the approximation. The statistical inference with respect to the parametric components via profile likelihood ratio test is also efficient. Generally speaking, the proposed simple estimation method for the partially linear regression model has two advantages that are worth noting. First, it greatly simplifies the computation burden in model estimation with little loss of efficiency. Second, it can be used to reduce the model bias by considering interaction between categorial predictor and continuous predictor, or between two continuous predictors in the nonparametric component of the model.

Though the partially linear regression model is a simple semiparameric model, the results have offered us more insights about the ``bias-efficiency'' tradeoff in semiparametric model estimations: when estimating the nonparametric components, pursing further bias reduction can increase the variance of nonparametric estimation, but it has little effect on the estimation of the parametric components of the model, and the efficient loss in the parametric part is small. Comparing to a much eased computational burden, such loss in efficiency in the parametric part can be negligible. Our study raised an interesting problem in semiparametric estimation: how to balance between the computation burden and the efficiency of the estimators while minimizing model bias. Our results can be generalized to estimate more broadly defined semiparametric models utilizing the partial consistency properties to fully exploit the information in the data.

\bibliography{cuixia}

\section{Appendix: assumptions and proofs}
\setcounter{equation}{0}
\renewcommand{\theequation}{A.\arabic{equation}}

We need the following conditions to prove our theoretical results:
\vspace{-0.5cm}
\begin{enumerate}[(a).]
\item $E|\varepsilon|^4<\infty$ and $E\|X\|^4<\infty$. \\[-1cm]
\item The support of the continuous component of $Z$ is bounded.\\[-1cm]
\item The functions $g(z^d, z^c)$, $\E(X|Z^d=z^d, Z^c=z^c)$, the density function of $Z$, and their corresponding second derivatives with respect to $z^c$ are all bounded. \\[-1cm]
\item $\Sigma$ is nonsingular.\\[-1cm]
\item In presence of discrete covariate in $Z$, assume that for any category, the number of samples lies in this category is large enough and of order $n$.
\end{enumerate}

For simplicity of presentation, we only
discuss the case of $Z=Z^c$ and prove Theorem \ref{asym-beta}. When $Z$ is of $2$-dimension, we mainly consider that one component of $Z$ is discrete or both components in $Z$ are highly correlated. For the former case, according to condition (e) it can be concluded that each category has a sample size of order $n$. So categories do not affect the following proof which leads to the results of Corollary \ref{asym-beta-case1} . For the latter case, assumption (\ref{closeness}) implies that the following proof can be easily generalized to obtain Corollary \ref{asym-beta-case2}. The proofs for both Corollary \ref{asym-beta-case1} and Corollary \ref{asym-beta-case2} are therefore omitted here.

{\bf\noindent Proof of Theorem \ref{asym-beta}.} First, based on standard operations in least squares estimation, we can obtain the decomposition
$\sqrt{n}(\hat{\beta}-\beta)=R_1+R_2$, where
\begin{equation}
\begin{split}
R_1 &=
\Big\{\frac{1}{n}\sum\limits_{j=1}^J\sum\limits_{i=1}^{I}\{X_{(j-1)I+i}-\frac{1}{I}\sum\limits_{i=1}^{I}X_{(j-1)I+i}\}^T
\{X_{(j-1)I+i}-\frac{1}{I}\sum\limits_{i=1}^{I}X_{(j-1)I+i}\}\Big\}^{-1}\\
&\hspace{0.1cm} \times
\Big\{\frac{1}{\sqrt{n}}\sum\limits_{j=1}^J\sum\limits_{j=1}^{J}\{X_{(j-1)I+i}-\frac{1}{I}\sum\limits_{i=1}^{I}X_{(j-1)I+i}\}^T
\{g(Z_{(j-1)I+i})-\frac{1}{I}\sum\limits_{i=1}^{I}g(Z_{(j-1)I+i})\}\Big\}\\
&\equiv{R_1^N}/{R_1^D}
\end{split}
\end{equation}
and
\begin{equation}
\begin{split}
R_2
&=
\Big\{\frac{1}{n}\sum\limits_{j=1}^J\sum\limits_{i=1}^{I}\{X_{(j-1)I+i}-\frac{1}{I}\sum\limits_{i=1}^{I}X_{(j-1)I+i}\}^T
\{X_{(j-1)I+i}-\frac{1}{I}\sum\limits_{i=1}^{I}X_{(j-1)I+i}\}\Big\}^{-1}\\
&\hspace{0.1cm} \times
\Big\{\frac{1}{\sqrt{n}}\sum\limits_{j=1}^J\sum\limits_{j=1}^{J}\{X_{(j-1)I+i}-\frac{1}{I}\sum\limits_{i=1}^{I}X_{(j-1)I+i}\}^T
\{\varepsilon_{(j-1)I+i}-\frac{1}{I}\sum\limits_{i=1}^{I}\varepsilon_{(j-1)I+i}\}\Big\}\\
&\equiv{R_2^N}/{R_2^D}\\
\end{split}
\end{equation}

Hereby we will show that the term $R_1$
converges to zero in probability as $n\rightarrow \infty$ and the
asymptotic distribution of $R_2$ is multivariate normal with zero
mean vector and covariance matrix given in (\ref{norm-beta}).

According to the form of $R_1$, we need to first analyze the numerator
$R_1^N$ and the denominator $R_1^D$ respectively. Let
$\mathcal{F}_n=\sigma\{Z_1,Z_2,\cdots,Z_n\}$ and observe that
conditionally on $\mathcal{F}_n$, $X_{(j-1)I+i}$ are independent of
each other. The following is a sketch.

We first analyze $R_1^N$. Denote $\E(X|Z=z)$ by $m(z)$ and $X-m(Z)$
by $e$, then
\begin{equation}
\label{R1N}
\begin{split}
R_1^N = &
\frac{1}{\sqrt{n}}\sum\limits_{j=1}^J\sum\limits_{i=1}^I\{m(Z_{(j-1)I+i})-\frac{1}{I}\sum\limits_{i=1}^Im(Z_{(j-1)I+i})\}
\{g(Z_{(j-1)I+i})-\frac{1}{I}\sum\limits_{i=1}^{I}g(Z_{(j-1)I+i})\}\\
&+\frac{1}{\sqrt{n}}\sum\limits_{j=1}^J\sum\limits_{i=1}^I\{e_{(j-1)I+i}-\frac{1}{I}\sum\limits_{i=1}^Ie_{(j-1)I+i}\}
\{g(Z_{(j-1)I+i})-\frac{1}{I}\sum\limits_{i=1}^{I}g(Z_{(j-1)I+i})\}\\
&=R_1^{N(1)}+R_1^{N(2)}.
\end{split}
\end{equation}

Notice that $R_1^{N(1)}$ can be expressed using the following summations,
\begin{equation*}
\begin{split}
&R_1^{N(1)} =
\frac{1}{\sqrt{n}I^2}\sum\limits_{j=1}^J\sum\limits_{i=1}^I\sum\limits_{l=1}^I\sum\limits_{k=1}^I
\{m(Z_{(j-1)I+i})-m(Z_{(j-1)I+l})\}
\{g(Z_{(j-1)I+i})-g(Z_{(j-1)I+k})\}\\
\end{split}
\end{equation*}
Parallel to the proof of \citet*{Hsing&Carroll:1992} and
\citet*{Zhu&Ng:1995}, we can show that
\begin{equation*}
\begin{split}
R_1^{N(1)}\leq&\frac{1}{\sqrt{n}I^2}\sqrt{\sum\limits_{j=1}^J\sum\limits_{i=1}^I\sum\limits_{l=1}^I\sum\limits_{k=1}^I
\|m(Z_{(j-1)I+i})-m(Z_{(j-1)I+l})\|^2}\\
&\hspace{0.5cm}\times\sqrt{\sum\limits_{j=1}^J\sum\limits_{i=1}^I\sum\limits_{l=1}^I\sum\limits_{k=1}^I
|g(Z_{(j-1)I+i})-g(Z_{(j-1)I+k})|^2}\\
=&O_P(n^{-1/2}I^{-2}n^{\delta})=o_P(1).
\end{split}
\end{equation*}
Here $\delta$ is a arbitrarily small positive constant. Let $\Omega_j$ denote the sample set lying in the $j$th partition with $1\leq j \leq J$.  The last equality obtained from the fact that, under condition (c),
$m(\cdot)$ and $g(\cdot)$ have a total variation of order $\delta$,
\begin{equation*}
\begin{split}
&\lim\limits_{n\rightarrow \infty} \frac{1}{n^{\delta}}
\sup\limits_{\{\Omega_j,1\leq j\leq J\}}\sum\limits_{i=1}^{I-1}
\|m(Z_{(j-1)I+i})-m(Z_{(j-1)I+(i+1)})\|=0,\\
&\lim\limits_{n\rightarrow \infty} \frac{1}{n^{\delta}}
\sup\limits_{\{\Omega_j,1\leq j\leq J\}}\sum\limits_{i=1}^{I-1}
|g(Z_{(j-1)I+i})-g(Z_{(j-1)I+(i+1)})|=0.
\end{split}
\end{equation*}

Next we consider $R_{1}^{N(2)}$. Let $\bar{e}_{(n)}$ and $\bar{e}_1$
be the largest and smallest of the corresponding
$e_i$'s, respectively. It is clear that
\begin{equation*}
\begin{split}
R_1^{N(2)}\leq &
\frac{\bar{e}_{(n)}-\bar{e}_1}{\sqrt{n}I}\sum\limits_{j=1}^J\sum\limits_{i=1}^I\sum\limits_{l=1}^I
|g(Z_{(j-1)I+i})-g(Z_{(j-1)I+l})|\\
=&2\frac{\bar{e}_{(n)}-\bar{e}_1}{\sqrt{n}I}\sum\limits_{j=1}^J\sum\limits_{1\leq
i<l\leq I}
|g(Z_{(j-1)I+i})-g(Z_{(j-1)I+l})|
\end{split}
\end{equation*}
The above argument leads to that
\begin{equation*}
\begin{split}
R_1^{N(2)}\leq&2\frac{\bar{e}_{(n)}-\bar{e}_1}{\sqrt{n}I}
\sum\limits_{i=1}^I\sum\limits_{l=1}^I\sum\limits_{j=1}^{n-1}
|g(Z_{(j+1)})-g(Z_{(j)})|\\
\leq&2I\frac{\bar{e}_{(n)}-\bar{e}_1}{\sqrt{n}}
\sum\limits_{j=1}^{n-1} |g(Z_{(j+1)})-g(Z_{(j)})|.
\end{split}
\end{equation*}
Applying Lemma A.1 of \citet*{Hsing&Carroll:1992}, we obtain
$$
n^{-1/4}|\bar{e}_{(n)}-\bar{e}_1|\stackrel{P}\longrightarrow 0.
$$
Note the fact that total variation of $g(\cdot)$ is of order
$n^{\delta}$, we have $R_1^{N(2)}=o_P(1)$. Combining the results
about $R_1^{N(1)}$ and $R_1^{N(2)}$, the proof for $R_1^N$ is
completed.

Next consider $R_1^D$ and $R_2^D$. Since $R_1^D=R_2^D$, we only need
to show the case of $R_1^D$. The expectation of $R_1^D$ is calculated as follows.
\begin{equation*}
\begin{split}
\E(R_1^D) = & \E(XX^{\top})-\frac{1}{nI}\sum\limits_{j=1}^J
\sum\limits_{i=1}^I\sum\limits_{l=1}^IE\{X_{(j-1)I+i}X_{(j-1)I+l}\}\\
=&\E(XX^{\top})-\frac{1}{nI}\sum\limits_{j=1}^J
\sum\limits_{i=1}^IE\{X_{(j-1)I+i}X_{(j-1)I+i}\}-\frac{1}{nI}\sum\limits_{j=1}^J
\sum\limits_{i\neq l}E\{X_{(j-1)I+i}X_{(j-1)I+l}\}\\
=&(1-\frac{1}{I})\E(XX^{\top})-\frac{1}{nI}\sum\limits_{j=1}^J
\sum\limits_{i\neq l}E\Big[E\{X_{(j-1)I+i}X_{(j-1)I+l}|\mathcal{F}_n\}\Big]\\
\end{split}
\end{equation*}
Under the assumption that conditionally on $\mathcal{F}_n$, $X_{(j-1)I+i}$ are independent of
each other, we can obtain that $E\{X_{(j-1)I+i}X_{(j-1)I+l}|\mathcal{F}_n\}=m(Z_{(j-1)I+i})m(Z_{(j-1)I+l})$. This, together with the above analysis, gives
\begin{equation*}
\begin{split}
\E(R_1^D) =&(1-\frac{1}{I})\E(XX^{\top})-\frac{I-1}{nI}\sum\limits_{j=1}^J
\sum\limits_{i= l}^IE\Big[m(Z_{(j-1)I+i})m(Z_{(j-1)I+i})\Big]\\
&-\frac{1}{nI}\sum\limits_{j=1}^J
\sum\limits_{i\neq l}E\Big[m(Z_{(j-1)I+i})\{m(Z_{(j-1)I+l})-m(Z_{(j-1)I+i})\}\Big]\\
=&(1-\frac{1}{I})\E(XX^{\top})-\frac{I-1}{nI}\sum\limits_{j=1}^J
\sum\limits_{i= l}^IE\Big[m(Z_{(j-1)I+i})m(Z_{(j-1)I+i})\Big]+o(1)\\
=&(1-\frac{1}{I})E\Big[\{X-\E(X|Z)\}\{X-\E(X|Z)\}^{\top}\Big]+o(1).
\end{split}
\end{equation*}
The term of order $o(1)$ is obtained following a similar argument of
Theorem 2.3 of \citet*{Hsing&Carroll:1992}. This completes the proof
for $R_1$.

We now deal with the term $R_2$. Observe that given $\{(X_i,Z_i),
i=1,\cdots,n\}$, each term of
$\{\varepsilon_{(j-1)I+i}-\frac{1}{J}\sum\limits_{j=1}^{J}\varepsilon_{(j-1)I+i}\}$
has mean zero and is independent of each other. Thus
$R_2$ is asymptotically normal with mean zero. We will show that the
limiting variance of $R_2$ is equal to the covariance matrix given
in (\ref{norm-beta}). That is,
\begin{equation*}
\begin{split}
\mbox{Var}(R_2|\{X_i,Z_i\}) = & (R_2^D)^{-1}\mbox{Var}(R_2^N|\{X_i,Z_i\})(R_2^D)^{-1}\\
=&
\{\E(R_2^D)\}^{-1}\E\{\mbox{Var}(R_2^N|\{X_i,Z_i\})\}\{\E(R_2^D)\}^{-1}+o_P(1)
\end{split}
\end{equation*}
and
\begin{equation*}
\begin{split}
\mbox{Var}(R_2^N|\{X_i,Z_i\}) = &
\frac{1}{n}\sum\limits_{j=1}^J\sum\limits_{i=1}^{I}\{X_{(j-1)I+i}-\frac{1}{I}\sum\limits_{i=1}^{I}X_{(j-1)I+i}\}
\{X_{(j-1)I+i}-\frac{1}{I}\sum\limits_{i=1}^{I}X_{(j-1)I+i}\}^{\top}\\
&\hspace{1cm}\times \E\Big[\{\varepsilon_{(j-1)I+i}-\frac{1}{I}\sum\limits_{i=1}^{I}\varepsilon_{(j-1)I+i}\}^2\Big|\{X_i,Z_i\}\Big]\\
=&\frac{\sigma^2}{n}\sum\limits_{j=1}^J\sum\limits_{i=1}^{I}\{X_{(j-1)I+i}-\frac{1}{I}\sum\limits_{i=1}^{I}X_{(j-1)I+i}\}
\{X_{(j-1)I+i}-\frac{1}{I}\sum\limits_{i=1}^{I}X_{(j-1)I+i}\}^{\top}\\
\stackrel{P}\longrightarrow &
\sigma^2(1-\frac{1}{I})\E\Big[\{X-\E(X|Z)\}\{X-\E(X|Z)\}^{\top}\Big].
\end{split}
\end{equation*}
Combining the last two equations, we complete the proof of
Theorem \ref{asym-beta}. $\Box$

{\bf\noindent Proof of Theorem \ref{asym-T1}.} First we show that $\mbox{RSS}_1=\sigma^2\{1+o_P(1)\}$. By (\ref{est}),
\begin{equation*}
\begin{split}
\mbox{RSS}_1&=\frac{1}{n}\sum\limits_{j=1}^J\sum\limits_{i=1}^I \{Y_{(j-1)I+i}-\hat{\alpha}_{j1}
-X_{(j-1)I+i}\hat{\bbeta}_{1}\}^2\\
&=\frac{1}{n}\sum\limits_{j=1}^J\sum\limits_{i=1}^I \Big[\{X_{(j-1)I+i}-\frac{1}{I}\sum\limits_{i=1}^{I}X_{(j-1)I+i}\}(\bbeta-\hat{\bbeta}_1)\\
&\hspace{2.5cm}+\{g(Z_{(j-1)I+i})-\frac{1}{I}\sum\limits_{i=1}^{I}g(Z_{(j-1)I+i})\}\\
&\hspace{2.5cm}+\{\varepsilon_{(j-1)I+i}-\frac{1}{I}\sum\limits_{i=1}^{I}\varepsilon_{(j-1)I+i}\}\Big]^2\\
&=I_1+I_2+I_3+I_4+I_5+I_6,
\end{split}
\end{equation*}
where
\begin{equation*}
\begin{split}
&I_1=\frac{1}{n}\sum\limits_{j=1}^J\sum\limits_{i=1}^I\{\varepsilon_{(j-1)I+i}-\frac{1}{I}\sum\limits_{i=1}^{I}\varepsilon_{(j-1)I+i}\}^2\\
&I_2=
\frac{1}{n}\sum\limits_{j=1}^J\sum\limits_{i=1}^I[\{X_{(j-1)I+i}-\frac{1}{I}\sum\limits_{i=1}^{I}X_{(j-1)I+i}\}
(\bbeta-\hat{\bbeta}_1)]^2\\
&I_3=\frac{1}{n}\sum\limits_{j=1}^J\sum\limits_{i=1}^I\{g(Z_{(j-1)I+i})-\frac{1}{I}\sum\limits_{i=1}^{I}g(Z_{(j-1)I+i})\}^2\\
&I_4=\frac{2}{n}\sum\limits_{j=1}^J\sum\limits_{i=1}^I\{\varepsilon_{(j-1)I+i}-\frac{1}{I}\sum\limits_{i=1}^{I}\varepsilon_{(j-1)I+i}\}
\{g(Z_{(j-1)I+i})-\frac{1}{I}\sum\limits_{i=1}^{I}g(Z_{(j-1)I+i})\}\\
&I_5=\frac{2}{n}\sum\limits_{j=1}^J\sum\limits_{i=1}^I\{X_{(j-1)I+i}-\frac{1}{I}\sum\limits_{i=1}^{I}X_{(j-1)I+i}\} (\bbeta-\hat{\bbeta}_1)\{g(Z_{(j-1)I+i})-\frac{1}{I}\sum\limits_{i=1}^{I}g(Z_{(j-1)I+i})\}\\
&I_6=\frac{2}{n}\sum\limits_{j=1}^J\sum\limits_{i=1}^I\{X_{(j-1)I+i}-\frac{1}{I}\sum\limits_{i=1}^{I}X_{(j-1)I+i}\} (\bbeta-\hat{\bbeta}_1) \{\varepsilon_{(j-1)I+i}-\frac{1}{I}\sum\limits_{i=1}^{I}\varepsilon_{(j-1)I+i}\}
\end{split}
\end{equation*}
Using the same arguments when analyzing $R_1$ and $R_2$, it can be shown that
\begin{equation*}
\begin{split}
&I_1=\frac{I-1}{I} \sigma^2\{1+o_P(1)\}, \hspace{1cm} I_2=O_P(n^{-1}),\hspace{1cm}I_3=o_P(n^{-1/2}),\\
&I_4=o_P(n^{-1/4}), \hspace{1cm} I_5=o_P(n^{-3/4}),\hspace{1cm}I_6=o_P(n^{-1/2}).
\end{split}
\end{equation*}
Similarly, $\mbox{RSS}_0$ can be decomposed as
\begin{equation*}
\begin{split}
\mbox{RSS}_0&=\frac{1}{n}\sum\limits_{j=1}^J\sum\limits_{i=1}^I \{Y_{(j-1)I+i}-\hat{\alpha}_{j0}
-X_{(j-1)I+i}\hat{\bbeta}_{0}\}^2\\
&=\frac{1}{n}\sum\limits_{j=1}^J\sum\limits_{i=1}^I \Big[\{Y_{(j-1)I+i}-\hat{\alpha}_{j1}
-X_{(j-1)I+i}\hat{\bbeta}_{1}\}\\
&\hspace{2cm}+\{X_{(j-1)I+i}-\frac{1}{I}\sum\limits_{i=1}^{I}X_{(j-1)I+i}\}(\hat{\bbeta}_{1}-\hat{\bbeta}_{0})\Big]^2\\
&=\mbox{RSS}_1+J_1+J_2,
\end{split}
\end{equation*}
with
\begin{equation*}
\begin{split}
&J_1=\frac{1}{n}\sum\limits_{j=1}^J\sum\limits_{i=1}^I\Big[\{X_{(j-1)I+i}-\frac{1}{I}\sum\limits_{i=1}^{I}X_{(j-1)I+i}\}(\hat{\bbeta}_{1}-\hat{\bbeta}_{0})\Big]^2\\
&J_2 = \frac{2}{n}\sum\limits_{j=1}^J\sum\limits_{i=1}^I \{Y_{(j-1)I+i}-\hat{\alpha}_{j1}
-X_{(j-1)I+i}\hat{\bbeta}_{1}\} \{X_{(j-1)I+i}-\frac{1}{I}\sum\limits_{i=1}^{I}X_{(j-1)I+i}\}(\hat{\bbeta}_{1}-\hat{\bbeta}_{0}).
\end{split}
\end{equation*}
From the proof of Theorem \ref{asym-beta}, it holds that $\frac{1}{n}\sum\limits_{j=1}^J\sum\limits_{i=1}^{I}\{X_{(j-1)I+i}-\frac{1}{I}\sum\limits_{i=1}^{I}X_{(j-1)I+i}\}
\{X_{(j-1)I+i}-\frac{1}{I}\sum\limits_{i=1}^{I}X_{(j-1)I+i}\}^{\top}\stackrel{P}\longrightarrow \frac{I-1}{I}\Sigma$. Furthermore, the estimators for $\bbeta$ under the null and alternative hypotheses then have the following relation
$$
\hat{\bbeta}_0=\hat{\bbeta}_1-\Sigma^{-1}A^\top\{A\Sigma^{-1}A^\top\}^{-1}A\hat{\bbeta}_1+o_P(\hat{\bbeta}_1).
$$
$J_1$ can then be written as
$$
J_1=\frac{I-1}{I}\hat{\bbeta}_1^{\top}A^\top\{A\Sigma^{-1}A^\top\}^{-1}A\hat{\bbeta}_1+o_P(\hat{\bbeta}_1).
$$
This, together with the asymptotic normality of $\hat{\bbeta}_1$ in Theorem \ref{asym-beta} implies that under the null hypothesis $A\hat{\bbeta}_1\stackrel{\mathcal{L}}\longrightarrow N(0,\sigma^2\frac{I-1}{I}A\Sigma^{-1}A^\top)$, we have
$
nJ_1\stackrel{\mathcal{L}}\longrightarrow \sigma^2 \chi^2_k.
$
By some calculation, it can be shown that $J_2=0$. Thus,
$$
n(\mbox{RSS}_0-\mbox{RSS}_1)\stackrel{\mathcal{L}}\longrightarrow \sigma^2 \chi^2_k.
$$
Then by Slutsky theorem,
$$
nT_n^1=n\frac{\mbox{RSS}_0-\mbox{RSS}_1}{\mbox{RSS}_1}\stackrel{\mathcal{L}}\longrightarrow  \frac{I}{I-1} \chi^2_k. \quad \quad \Box
$$

\end{document}